\documentclass[12pt,preprint]{aastex}

\usepackage{verbatim}
\usepackage{rotating}
\usepackage{lscape}

\def\h2{$\rm H_2$}

\def\solmass{$\rm M_{\sun}$}

\newcommand{\msun}{M$_{\odot}$}
\newcommand{\halpha}{H$\alpha$}
\newcommand{\hii}{H{\sc II}}
\newcommand{\hi}{H{\sc I}}

\newcommand{\hoi}{Ho~{\sc I}}
\newcommand{\hoii}{Ho~{\sc II}}
\newcommand{\hoix}{Ho~{\sc IX}}
\newcommand{\mh}{$[$$\frac{M}{H}$$]$}

\title{The Recent Star Formation Histories of M81 Group Dwarf Irregular Galaxies}
\slugcomment{{\sc Accepted to ApJ:} July 31st, 2008}
\author{Daniel R.\ Weisz, Evan D.\ Skillman}
\affil{Astronomy Department, University of Minnesota,
Minneapolis, MN 55455}
\email{dweisz@astro.umn.edu, skillman@astro.umn.edu}

\author{John M.\ Cannon}
\affil{Department of Physics \& Astronomy, Macalester College, 1600 Grand Avenue, St. Paul, MN 55125}
\email{jcannon@macalester.edu}

\author{Andrew E.\ Dolphin}
\affil{Raytheon Corporation,
1151 E Hermans Rd, Tucson, AZ 85706}
\email{adolphin@raytheon.com}

\author{Robert C.\ Kennicutt, Jr.}
\affil{Institute of Astronomy, University of Cambridge, Madingley Road, Cambridge CB3 0HA, UK}
\email{robk@ast.cam.ac.uk}

\author{Janice Lee}
\affil{Observatories of the Carnegie Institution of Washington,
813 Santa Barbara Street, Pasadena, CA 91101}
\email{jlee@ociw.edu}

\author{Fabian Walter}
\affil{Max-Planck-Institut f\"{u}r Astronomie, K\"{o}nigstuhl 17, D-69117 Heidelberg, Germany}
\email{walter@mpia.de}

\author{}
\affil{}
\email{}

\clearpage

\begin{abstract}
We present observations and analysis of nine dwarf irregular galaxies (dIs)  in the 
M81 Group taken with the Advanced Camera for Surveys aboard the Hubble 
Space Telescope.  The nine galaxy sample (the Garland, M81 Dwarf~A, DDO 53, 
\hoix, \hoi, DDO 165, NGC 2366, \hoii, and IC 2574) spans
6 magnitudes in luminosity, a factor of 1000 in current star 
formation rate, and 0.5 dex in metallicity
allowing us to study star formation (SF) under a variety of different conditions.  
Here we use color-magnitude diagrams of resolved stellar populations to study the
star formation histories (SFHs) of these galaxies. 
Based on luminosity we divide the sample into faint and bright galaxies, with a dividing line of M$_{B}$ $=$ $-$15, and then analyze the similarities and 
differences in the SFHs, birthrate parameters, fraction of stars formed per time interval, and spatial distribution of stellar components.  Comparing these parameters as a function of luminosity, we find only minor differences in SF characteristics among the M81 Group dIs despite a wide range of physical properties.  We extend our comparison to select dIs in the Local Group (LG), with similar quality photometry,  and again find only minor differences in SF parameters.  The lack of a clear trend in SF parameters over a wide range of diverse environments suggests that SF in low mass systems may be dominated by stochastic processes.   The fraction of stars formed per time interval for an average M81 Group and LG dI is consistent with a constant SFH.  However, individual galaxies can show significant departures from a constant SFH. Thus, we find this result underlines the importance of stochastic SF in dIs.
In addition to a statistical comparison, we compare possible formation scenarios of the less luminous and candidate tidal dwarfs in the M81 Group.  The SFHs and the lack of an overdensity of associated red stars suggest that the Garland and \hoix\ are not dIs and are potentially tidal dwarf galaxies.  Interestingly, a noteworthy difference between the LG and the M81 Group is the lack of tidal dwarf candidates in the LG.

\end{abstract}

\keywords{
galaxies: dwarf ---
galaxies: stellar content ---
galaxies: formation ---
galaxies: evolution
}

\begin{document}

\section{Introduction}

\subsection{The M81 Group Dwarf Galaxy Project}

The ability to translate resolved stellar populations into star formation 
histories (SFHs) has become an increasingly powerful method for understanding 
how star formation (SF) shapes galaxy evolution \citep[e.g.,][]{gza05}.   
The power of the Hubble Space Telescope (HST) is truly being realized as we 
are able to resolve the stellar components of increasingly distant galaxies. With the continuing expansion of the sample of galaxies with measured 
SFHs from resolved stars \citep[e.g., the ACS Nearby Galaxy Survey Treasury (ANGST),][]{dal08}, we are able to study how a wide variety of galaxies 
evolve both individually and in their respective environments.

Of particular interest are star forming galaxies that sample the faint end of the galactic luminosity function.  Characteristics of SF in this regime are not as well studied as brighter, more massive galaxies, due to completeness limits of large galaxy surveys (e.g., SDSS). For example, \citet{lee07} and \citet{ken08} empirically find an increased dispersion in the \halpha\ equivalent width (EW) measurements, the normalized SFR over the past 5 Myr, of star forming galaxies fainter than $M_{B}$ $\sim$ $-$15 compared to brighter galaxies.  The cause behind the increased dispersion is not clear, and we could simply be observing the galaxies at different points in their SF duty cycle.  Alternatively, different physical mechanisms could be responsible for the observed dispersion. There are several candidate theories that have distinct signatures: stochastic effects \citep[cf.,][]{mue76, ger78}, internal feedback \citep[stellar winds, supernovae, e.g.,][]{pel04, sti07}, external interactions \citep[mergers, tidal influence, e.g.,][]{tom72, gne99}, or a combination thereof.  Direct observational evidence of recent SF episodes can shed light into the role of each process.

Specifically, dwarf irregular galaxies (dIs) are excellent environments for probing the mechanisms which regulate SF.  By number, dIs dominate the faint star forming galaxy count and span a wide range in mass, luminosity, metallicity, and SFR \citep{mat98}.  They are solid body rotators \citep{ski88}, so that features resulting from past SF episodes are preserved and not destroyed as they are in larger galaxies with differential rotation \citep{ski96}.  Thus, SFHs based on resolved stellar populations of dIs directly probe the physical mechanisms that govern SF in low mass galaxies.  

Additionally, dIs are useful for understanding galaxy interactions within a group environment.  Because SFHs trace the evolution of each galaxy over the history of the universe, we can measure the fraction of stellar mass formed at different epochs, which is directly comparable to models of galaxy interactions within groups \citep[e.g.,][]{ric05, orb08}.  At more recent times, SFHs of dIs can be used to quantify patterns and duty cycles of SF over the past $\sim$ 1 Gyr and compare them to theoretical predictions and amongst different groups of galaxies.

Comparing nearby galaxy groups, an interesting feature of the M81 Group is the presence of suspected tidal dwarfs \citep[e.g.,][]{mak02, kar04}, which have no counterparts in the Local Group.  A three body interaction between larger galaxies in the M81 Group (M81, M82, and NGC 3077) occurred $\sim$ 300 Myr ago, and there have been numerous observations of \hi\ tidal debris leftover from this event \citep[e.g.,][]{van79,yun94}.  Young stellar systems have been observed in these tidal remnants, and we may be witnessing the formation of new dwarf galaxies (i.e., tidal dwarfs) in the nearby universe \citep[e.g.,][]{wal99b,mak02}.  

In this paper, we introduce the observations, color-magnitude diagrams (CMDs), and SFHs of nine dIs in the M81 Group based on HST/ACS imaging.  While our focus is on the recent temporal SFHs, we can also place loose constraints on the ancient SFHs and use red and blue star density maps to trace locations of past SF episodes.  Using the recent SFHs, we are able to quantify and compare the strength and duration of SF episodes amongst the sample and to similarly observed galaxies in the LG.  We use this paper to establish analysis techniques that we anticipate applying to a larger sample in the future.  Here, we present a limited comparison of the SFHs to observations in other wavelength regimes (e.g., \halpha , ultraviolet (UV), infrared (IR), \hi), and in our future work we will compare the spatially resolved recent SFHs to \hi\ observations and calibrate different SFR indicators to our CMD based SFHs.

\subsection{The M81 Group Dwarf Galaxy Sample}

The M81 Group presents an excellent environment for the study of resolved stellar populations and SFHs.  The M81 Group is a prominent group of galaxies close enough to the Local Group \citep[with distances typically $\sim$ 3.8 Mpc,][]{kar02} that the HST/ACS easily resolves the bright, young stellar populations.  Further, the kinematic state of the M81 Group is very dynamic.  \hi\ observations and related models of the M81 Group \citep{yun94, yun99} reveal a recent ($\sim$ 300 Myr) interaction between the major member galaxies and the presence of tidal streams and debris.  A three dimensional map as well as detailed structural and kinematical information of the M81 Group are presented in \cite{kar02}. 

The sample of nine M81 Group dIs was selected from galaxies which had excellent ancillary observations, starting with galaxies observed as part of the Spitzer Infrared Nearby Galaxies Survey \citep[SINGS;][]{ken03}.
There are eight M81 Group dIs in both the SINGS and The \hi\ Nearby Galaxy Survey \citep[THINGS;][]{wal08}.
This list was augmented by adding NGC 2366 (which had similar quality ancillary 
observations, although not part of the SINGS survey), the Garland (for which 
HST/ACS observations already existed, GO-9381, PI: Walter), and UGC 4483,
which is a dwarf starburst galaxy \citep[cf.,][]{sea94, vss98}.  Two galaxies were subsequently dropped from the sample.
UGC 4483 was observed on December 31, 2006,
but the observation failed; this was shortly followed by the failure of the 
ACS, so we have no new observations for it and it was dropped from the sample.  M81 Dwarf~B was dropped from the sample after  we found it to be significantly 
more distant than the other galaxies in agreement with other recent analysis \citep[e.g., 5.3 Mpc,][]{kar02}.  The final sample of nine M81 Group dIs spans a wide range of properties including 6 magnitudes in luminosity, a factor of 1000 in current star formation rate (SFR), and 0.5 dex in metallicity (see Table \ref{tab1}).   

\begin{figure}[t]
\begin{center}
\plotone{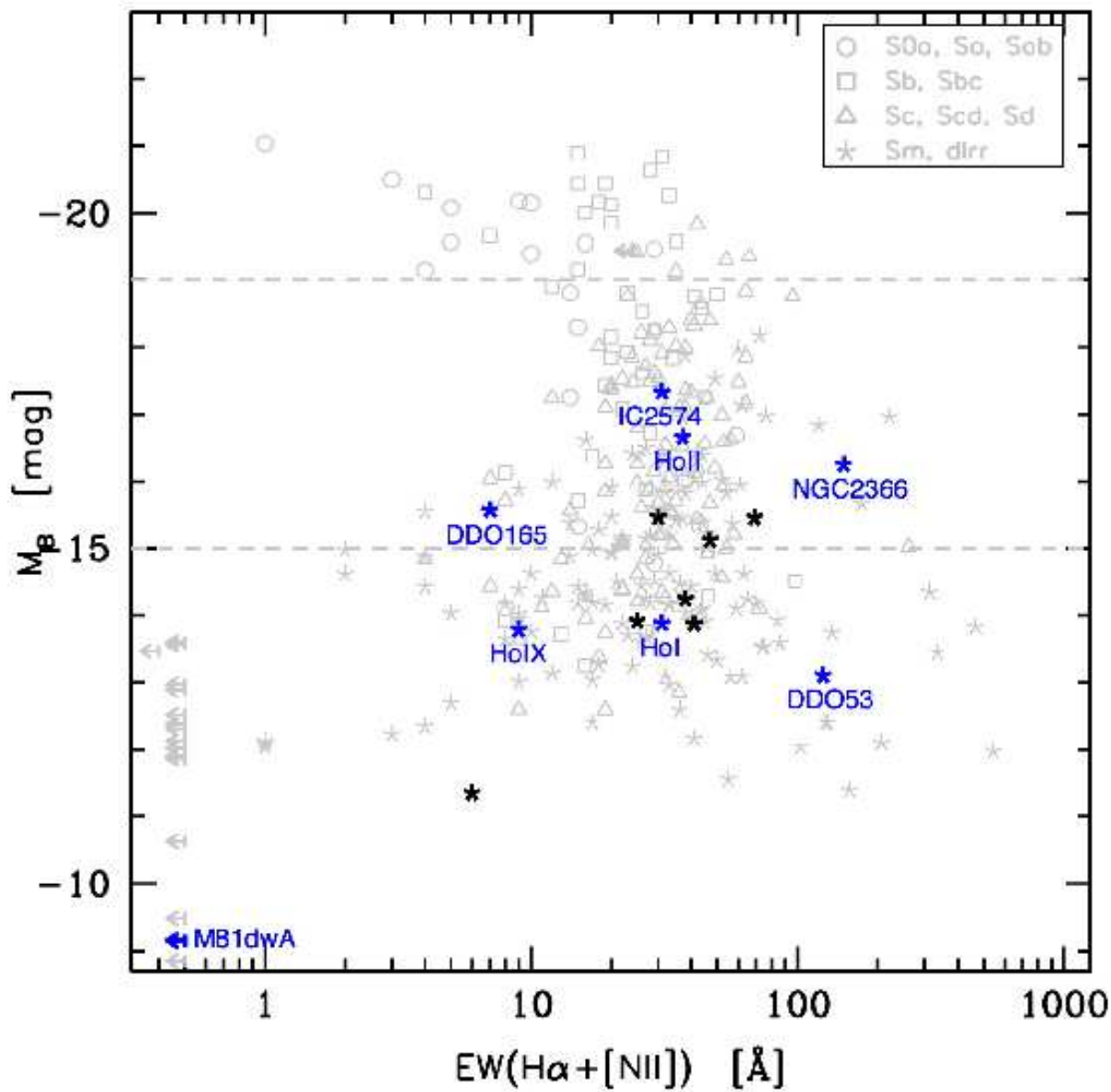}
\caption{The Local Volume star-forming galaxy sequence plotted as as \halpha\ EW vs. M$_{B}$ \citep{lee07,ken08}.  The M81 Group dIs in this paper are shown in blue, and the Local Group dIs in this paper are shown in black.  Note the wide range of luminosities and current SFRs (\halpha\ EW) that the dIs in our sample span. }  
\label{ha_ew}
\end{center}
\end{figure}

In Figure \ref{ha_ew} we have highlighted the dIs from this paper (M81 Group dIs in blue and LG dIs in black) in the context of the Local Volume star-forming galaxy sample \citep{lee07, ken08}.   The galaxies here are shown in the M$_{B}$$-$\halpha\ equivalent width (EW) plane, which traces mass and current normalized SFR, respectively.  \citet{lee07} find an empirical transition at M$_{B}$ $\sim$ $-$15 based on the dispersions in \halpha\ EW. Our sample of dIs straddles both sides of this division, allowing us to probe differences in SF characteristics in both luminosity regimes over a wide range of current SFRs. 

Within our sample, we have two candidate tidal dwarf galaxies, the Garland and \hoix.  We will adopt that label throughout this paper, recognizing that the status of these features as true galaxies is controversial \citep[e.g.,][]{vand98, mak02, kar02} and they may simply be stellar systems forming in the outskirts of their larger companions (see Figure \ref{hi_map}).  We chose to exclude the candidate tidal dwarfs from our statistical analysis because of we find them to be fundamentally different than dIs in the sample.

\section{Observations and Photometry}

We obtained HST/ACS  \citep[][]{for98} images of seven of the nine M81 Group dIs with ACS/WFC between January 27, 2006 and December 30, 2006 (GO-10605) using the filter combination of F555W (V) and F814W (I) in order to optimize the combination of photometric depth and temperature baseline (represented by stellar color in the observation plane).  The requisite photometric depth of a signal to noise ratio of $\sim$ 5 at M$_{V} = -0.5$ allows us to sufficiently resolve CMD features that give us leverage on the recent SFHs \citep{dom02}.  The selection of filters is important because a larger difference in the central wavelengths of two selected filters provides for a clearer separation between features in the CMD. While observing at very blue and red wavelengths (e.g., U or B and I) would provide for a greater color difference, it also makes the integration time to the same photometric depth increase greatly.  Thus, to optimize both color separation and photometric depth we observed the M81 Group dIs using F555W and F814W.

For most of the galaxies, a single ACS field was sufficient to cover a significant fraction of the stars, for the larger galaxies, NGC 2366 and \hoii, we observed two fields each, and  
for IC 2574, an ACS field from the HST Archive was 
supplemented by two additional fields.  
The ninth galaxy, the Garland, was also added from the HST archive.  Both of these galaxies were observed in the same filter combination for comparable integration times.  All observations were CR-split to reduce the impact of cosmic rays and dithered to cover the chip gap. Most of the M81 Group is also in the HST's continuous viewing zone, making for very efficient observing
strategies \citep[e.g.,][]{cads98}. 

After the images were processed with the HST pipeline, we performed photometry using DOLPHOT, a version of HSTphot \citep{dol00} optimized for ACS observations.  Cosmic rays, hot pixel residuals, and extended objects were all rejected based on their brightness profiles.  Remaining objects were then subject to further photometric tests (signal to noise, crowding, and sharpness) resulting in catalogs of only well-measured stars (signal-to-noise $>$ 5, $| F555W_{sharp} + F814W_{sharp} | > 0.27$, and $ F555W_{crowd} + F814W_{crowd}  < 1.0$), where sharpness and crowding follow the defitions in \citet{dol00}. Artificial star tests were performed to determine the completeness limit for each field and these values are listed in Table \ref{tab2}.

The faintest well measured stars in our photometry have M$_{V}$ $\sim$ 0, which is adequate for recent SFHs, but only allows us to place weak constraints on the ancient ($>$ 6 Gyr) SFHs. The main restriction resulting from shallower CMDs is the loss of time resolution at older ages \citep{dol02}.
For example, the study of the Local Group galaxy Leo A by \citet{tol98} used
similar depth photometry (M$_{V}$ $\sim$ 0) to infer that bulk of the SF in Leo A occurred at later (more recent) times.  This conclusion was verified by the very deep photometry, 
including the oldest MS turnoffs, presented by  \citet{col07}.  While the emphasis of our work is to constrain the recent SFHs, for which we have sufficient photometric depth \citep[e.g.,][]{dea97}, we can place loose constraints on the ancient SFHs by placing all SF $>$ 6 Gyr into a single old age time bin.

\section{Methodology}

\subsection{Features on the CMD}

\begin{figure}[t]
\begin{center}
\epsscale{0.9}
\plotone{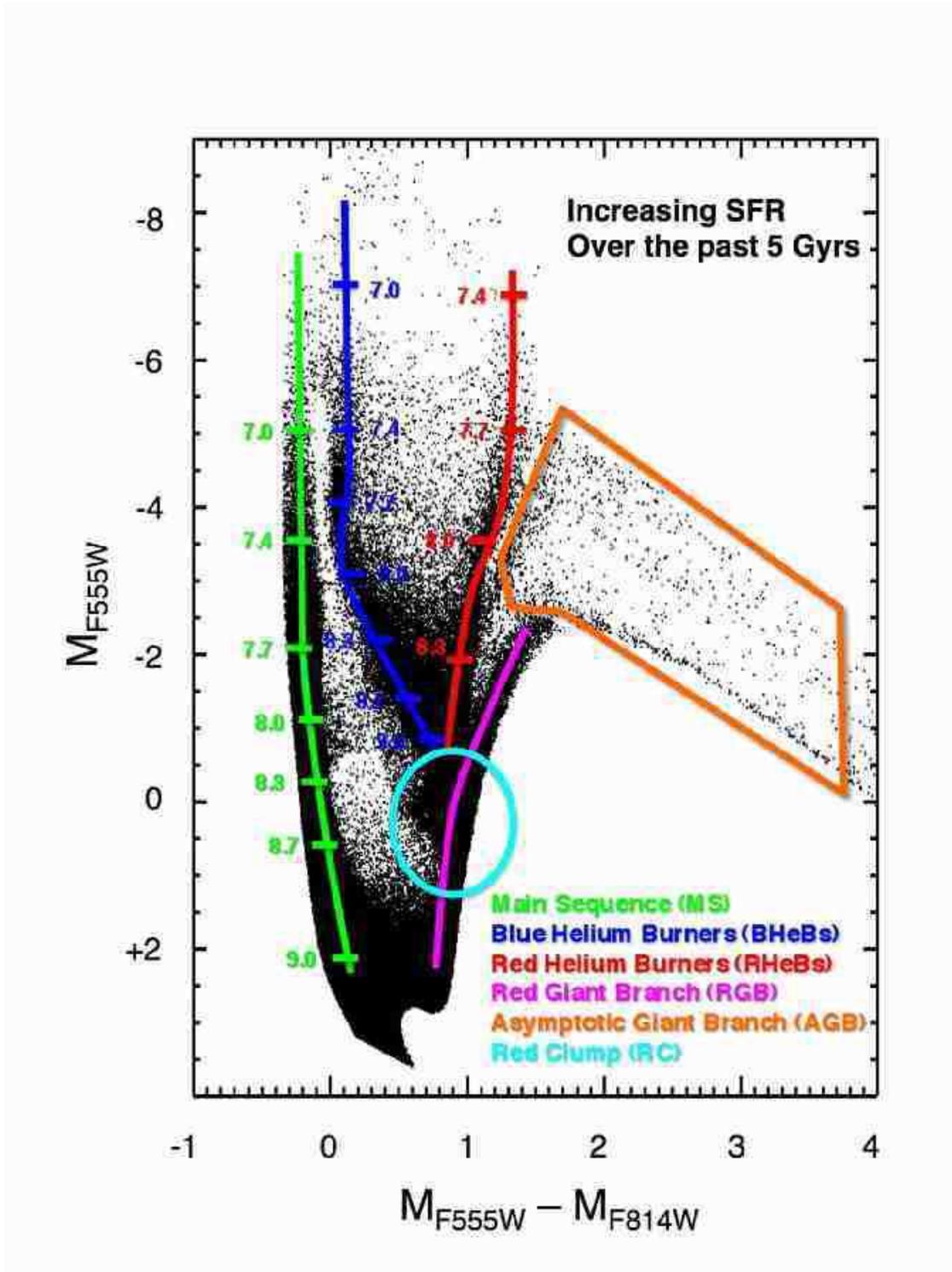}
\caption{\footnotesize{A simulated CMD of increasing SF over the past 5 Gyr.  Important features on the CMD are highlighted: Main Sequence (MS, green), Blue Helium Burners (BHeBs, blue), Red Helium Burners (BHeBs, red), Asymptotic Giant Branch (AGB, orange), Red Giant Branch (RGB, magenta), and Red Clump (RC, cyan).  For the MS, BHeBs, and RHeBs  we overlay the logarithmic ages of stars at a given magnitude \citep{mar07}.}}
\label{cartoon_cmd}
\end{center}
\end{figure}

In order to highlight important features on the CMD, we present a simulated CMD (Figure \ref{cartoon_cmd}) made with the stellar evolution models of \citet{mar07}. We have also overplotted the timescales from the theoretical isochrones on the young stellar sequences to demonstrate the relationship between age and magnitude. Here, we provide a brief overview of main features of the CMD.  For a more thorough review see \citet{gza05}.  The main sequence (MS) stars burning hydrogen in their cores are marked by the green line.  Immediately to the right of the MS are the blue helium burning stars (BHeBs), indicated by the blue line. These are intermediate mass stars ($\sim$ 2 $-$ 15 \msun) that have evolved off the MS and are burning helium in their cores. In a parallel track to the right of the BHeBs are the red helium burning stars (RHeBs), denoted by the red line. The HeB phase of stellar evolution is very short lived, making the HeB stars excellent chronometers for SF episodes as discussed in \S 3.2.  Note the BHeBs are roughly 2 magnitudes brighter than their corresponding MS stars of the same age.  In the orange box to the right of the RHeBs are asymptotic giant branch stars (AGBs) which are low to intermediate mass evolved stars burning both hydrogen and helium in shells.  The magenta line marks the red giant branch stars (RGBs), which are low to intermediate mass stars that have evolved off the MS and are burning hydrogen in shells surrounding the cores.  The tip of the red giant branch (TRGB), where the helium core of RGBs ignite, is an excellent distance indicator as it has a virtually invariant absolute I magnitude at low metallicities \citep{lee93}.  The red clump (RC) is a phase where low and intermediate mass stars lie on the CMD as they burn helium in their cores and is enclosed by the cyan circle.  Note that the RC overlaps with RGB stars and also merges with the BHeBs at (at $\sim$ 1 Gyr old BHeBs) and RHeBs (at $\sim$ 500 Myr for RHeBs) on the bright side, making this a very dense region on the CMD.

\subsection{Connecting CMDs to Recent SFHs}

\begin{figure}[t]
\begin{center}
\epsscale{0.95}
\plotone{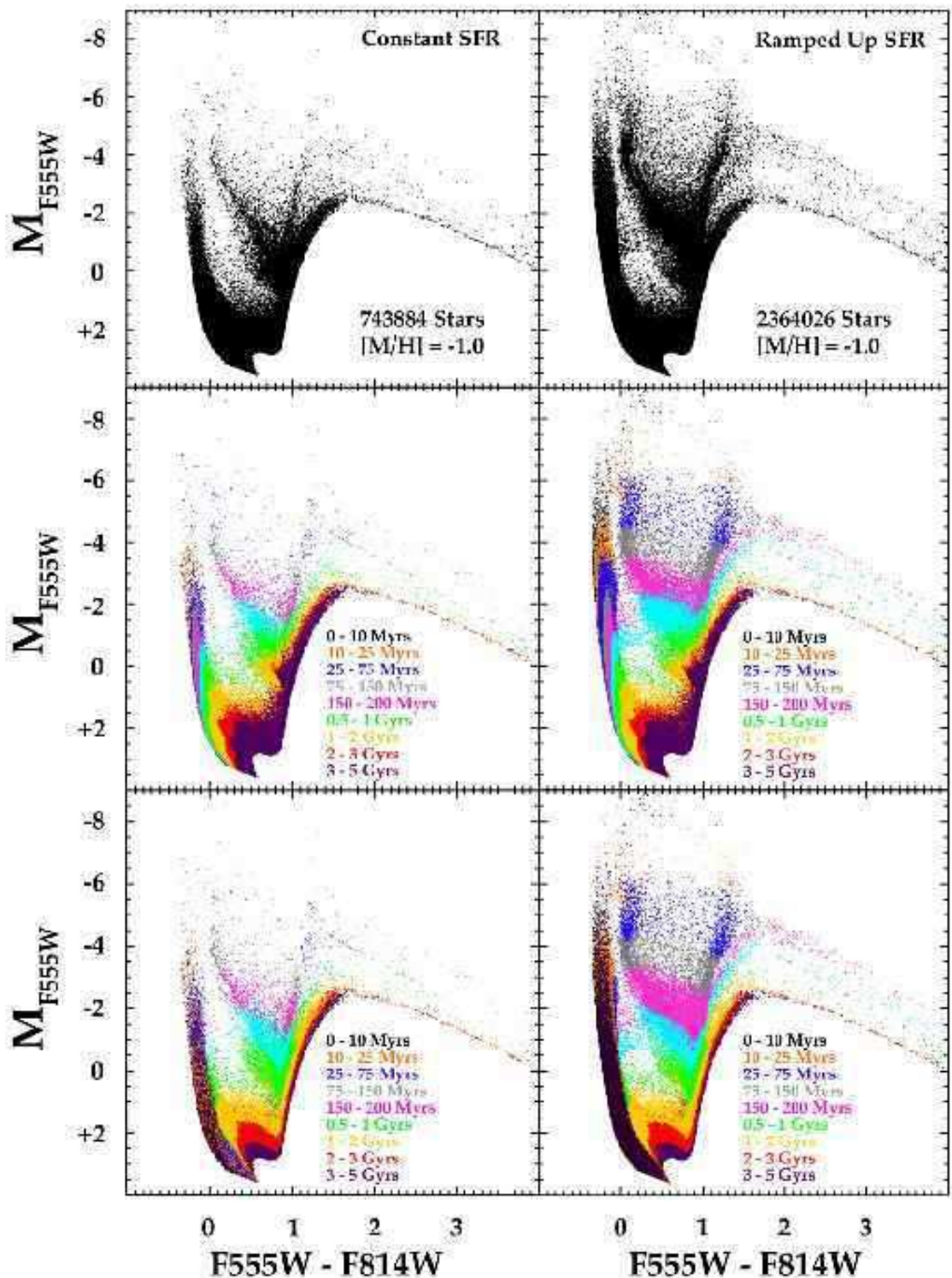}
\caption{\scriptsize{Simulated CMDs using models of \citet{mar07} that demonstrate the power of HeBs as chronometers of SF within the last $\sim$ 1 Gyr.  Panels on the left were constructed using a constant SFR (0.005 \msun\ yr$^{-1}$), while panels on the right represent a ramped up SFR (0.005 $-$ 0.1 \msun\ yr$^{-1}$).  The stars are color coded by age and middle and bottom panels are plotted in opposite order.  The points are plotted in reverse order to demonstrate the physical reality that HeBs, particularly BHeBs, of different ages do not have overlapping magnitudes and colors, making them very accurate chronometers for recent ($<$ 1 Gyr) SF.  In contrast, MS stars of different ages can overlap in magnitude and color, which provide limited leverage on SF over the past $\sim$ 1 Gyr.}}
\label{diag_cmd}
\end{center}
\end{figure}

An important step in verifying the accuracy of any 
derived SFH is being able to relate features of a given CMD to the corresponding SFH.
We have constructed an instructional aid by simulating two CMDs from simple input 
SFHs and color-coding the ages of the stars (see Figure \ref{diag_cmd}).  
Similar diagrams can be found in the literature \citep[e.g.,][]{gza05}, but 
here we emphasize the \textit{recent} SF by
limiting our input SFH to the last 5 Gyr.
These CMDs were created with the code of \citet{dol02} and stellar evolution 
models of \citet{mar07} using a power law initial mass function (IMF) with a spectral index of $\gamma$ $=$ $-1.30$ \citep[i.e., $\gamma$ $\propto$ $\frac{dn}{d(\log(m))}$, where $\gamma$ $=$ $-1.35$ represents the Salpeter IMF,][]{sal55}, between 0.15 and 100 \solmass, a binary fraction (BF) of 0.35, zero foreground reddening, and a constant metallicity (of a typical dI) of $[M/H]$ $=$ $-1.0$ (i.e., Z $=$ $0.1$ $Z_{\odot}$ and where $[M/H]$ is the logarithmic relative abundance of heavier elements, M, with respect to hydrogen, H, relative to the solar value).  

The CMDs in the left column of Figure \ref{diag_cmd} were created assuming a constant SFR of 0.005 \solmass~yr$^{-1}$ (typical for a dI) for the past 5 Gyr.   To more clearly demonstrate the positions of younger stars on the CMD, the CMDs in the right column were created with an increasing SFR over the past 5 Gyr, such that SFR$_{5Gyr}$ $=$ 0.005 \solmass~yr$^{-1}$ and SFR$_{Present}$ $=$ 0.1 \solmass~yr$^{-1}$. Because our emphasis is primarily on recent SF, we will not analyze the contributions of   AGB and RGB stars because they are indicators of SF older than $\sim$ 1 Gyr.  

In the top two panels of Figure \ref{diag_cmd} we present the CMDs as they would appear as if we were to observe this simulated galaxy.  We present the same CMDs  with the stars color-coded by ages in the lower four panels.  

In the center panels of Figure \ref{diag_cmd}, we present the constant SFR (left panel) and increasing SFR (right panel) CMDs with color coded ages.  Evident in both panels are the distinct age divisions of the MS and BHeB stars as a function of  magnitude.  That is, the youngest stars (0 $-$ 10 Myr, black points) are located at the brightest magnitudes and older stars are located at fainter magnitudes (i.e., 10 $-$ 25 Myr, orange points; 25 $-$ 75 Myr, blue points; 75 $-$ 150 Myr, grey points, etc) and they are divided into fairly distinct age divisions.  The MS shows these divisions all the way back to 5 Gyr, the oldest age of these CMDs, while the BHeBs extend back to $\sim$ 1 Gyr, where they merge into the RC.  The RHeBs also demonstrate a similar magnitude-age relationship, but blend into the RC at a brighter magnitude, or younger age.  

Considering only the MS in Figure \ref{diag_cmd}, we see stars of different ages can have the same magnitude and color.  For example, while the magenta points (150 $-$ 300 Myr) appear as a distinct age division on the MS, this is only a stripe of the brightest magenta MS colored stars. The fainter magenta colored MS stars are obscured on this CMD by older, fainter MS stars. Even stars of ages younger than 150 Myr (0 $-$ 10 Myr, black points; 10 $-$ 25 Myr, orange points; 25 $-$ 75 Myr, blue points; 75 $-$ 150 Myr, grey points) can have the same color and magnitude as the magenta colored stars.  To further emphasize this important degeneracy, in the lower panels we plot the same CMDs, only with the plotting order of the points reversed (young on top of old vs. old on top of young).  The reversal of the plotting order demonstrates that stars with the same magnitude and color on the MS can have vastly different ages such that the youngest stars (0 $-$ 10 Myr, black points) populate the entire MS to very faint magnitudes, and can occupy the same CMD space as 5 Gyr old MS stars.

In contrast, the BHeBs are particularly powerful chronometers for recent SF episodes because BHeBs of different ages have distinct magnitudes and colors.  Consider the magenta points (150 $-$ 300 Myr) on the BHeB sequence in the two center panels of Figure \ref{diag_cmd}.  There are no stars of any other age that have the same color and magnitude as the magenta colored BHeB stars.  This fact is further accentuated by the reverse plotted CMDs in the lower two panels.  Whereas the MS stars do not retain the same neat age divisions between the center and lower panels, the BHeBs do.  A slight caveat applies to the boundaries between the different generations of BHeBs.  There is a slight overlap of different age stars in these regions, but this effect is on order of 10\% according to simulations we have conducted.  Thus, the BHeBs provide us with extremely powerful leverage to accurately measure the recent SFH of a galaxy.


\subsection{Method of Measuring SFHs}

Technological and theoretical advances in the past decade have made it possible 
to derive SFHs of stellar populations with unprecedented detail and accuracy.  
Sophisticated programs \citep{tgf89, ts96, gabc96, m97, hol99, hvg99, dol02, 
ia02, yl07}  
enable us to directly compare observed stellar populations with model stellar 
populations and measure the corresponding SFH.  

Explicitly, a SFH is the SFR as a function of time and metallicity (i.e., SFR(t,Z)).  To derive the SFH of each M81 Group dI in our sample, we used the maximum likelihood method of \citet{dol02}.  The program creates synthetic CMDs using the models of \citet{mar07} for every combination of fixed (IMF, BF, etc.) and searchable (distance, extinction, metallicity, and SFR) variables.  The synthetic CMDs are then compared to the observed CMD using a modified $\chi^2$ parameter.  The best fit synthetic CMD yields the most likely SFH of the galaxy.  See \citet{dol02} for a full description of the SFH measurement algorithm.  

Along with the SFR, metallicity information is extracted from the best fit isochrone at each age. These fits yield a mean value of $[M/H]$ in each time bin allowing us to trace the chemical evolution of a galaxy.  Because our photometry does not reach the ancient MS turnoff, it is hard to place tight constraints on the chemical evolution from CMD fitting alone.  To aid in our solutions, we require that the program fit monotonically increasing metallicities. For deeper CMDs this assumption would not be necessary.

To ensure that we have a robust SFH, we compute both systematic and statistical uncertainties of the SFR in each time bin.  The systematic uncertainties are a product of the stellar evolution isochrones and are manifested as a degeneracy between distance and extinction. To account for this form of uncertainty, we re-solve for new SFHs allowing the distance and extinction values to vary slightly from the values associated with the most likely SFH.   The uncertainties in the SFRs for each variation are added in quadrature to produce the systematic component of our uncertanties.  The statistical uncertainties are a result of counting statistics in each CMD bin.  We use Monte Carlo tests to re-solve for SFHs after redistributing the stars in each bin according to a Poisson distribution.  The resultant systematic and statistical uncertainties are added in quadrature to produce the error bars in our final SFHs.  A more detailed discussion of the calculation and interpretation of uncertainties in SFHs can be found in \citet{dol02}.

\section{CMDs, SFHs, and Stellar Spatial Distributions}

The three main tools for our analysis are CMDs, SFHs, and stellar density maps for blue and red stars.  The combination of these three products aids in our analysis of the temporal and spatial properties of the SFHs for each galaxy.  In this section, we present the CMDs of each ACS observation, the lifetime and recent SFHs of each galaxy, and the stellar density maps for each galaxy.  As a setup for our later comparison of SF based on luminosity, the sample is divided into two luminosity regimes with a dividing line of M$_{B}$ $=$ $-$ 15.

\begin{figure}[t]
\begin{center}
\epsscale{0.9}
\plotone{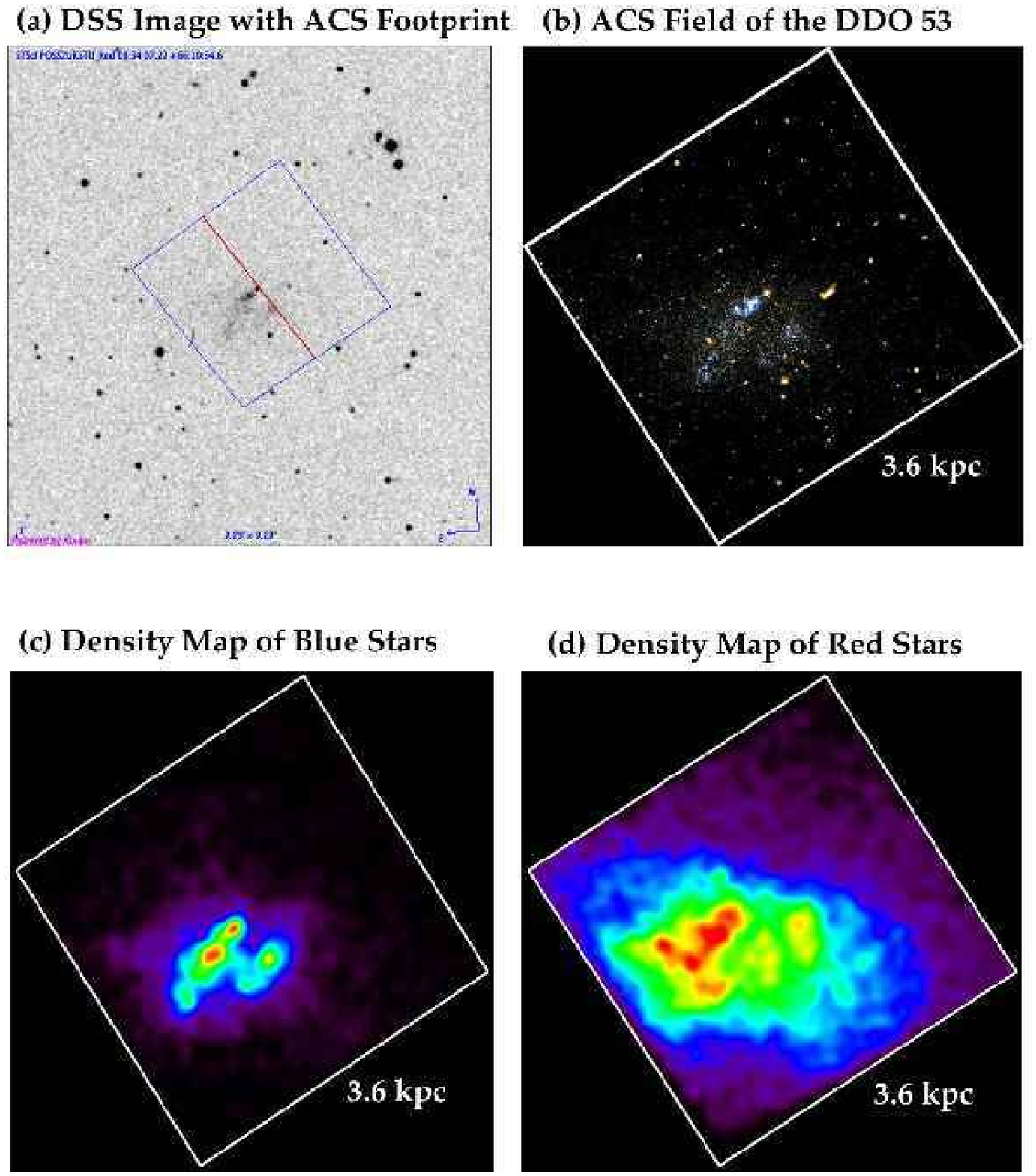}
\caption{A four panel figure of the DDO 53 showing the ACS footprint in blue with the chip gap direction in red over the R-Band DSS image (a), the color ACS image (b), and stellar density maps of blue (c) and red stars (d) as determined by their positions on the CMD.  The blue stars have a color $<$ 0.6 and magnitude $>$ 60\% completeness in F814W.  The red stars are defined as 0.6 $\leq$ color $\leq$ 3.0 and a magnitude range between the TRGB and 60\% completeness in F814W.  The highest density regions are the red contours and the lowest are in purple.}
\label{ddo53_image}
\end{center}
\end{figure}

\begin{figure}[t]
\begin{center}
\epsscale{0.9}
\plotone{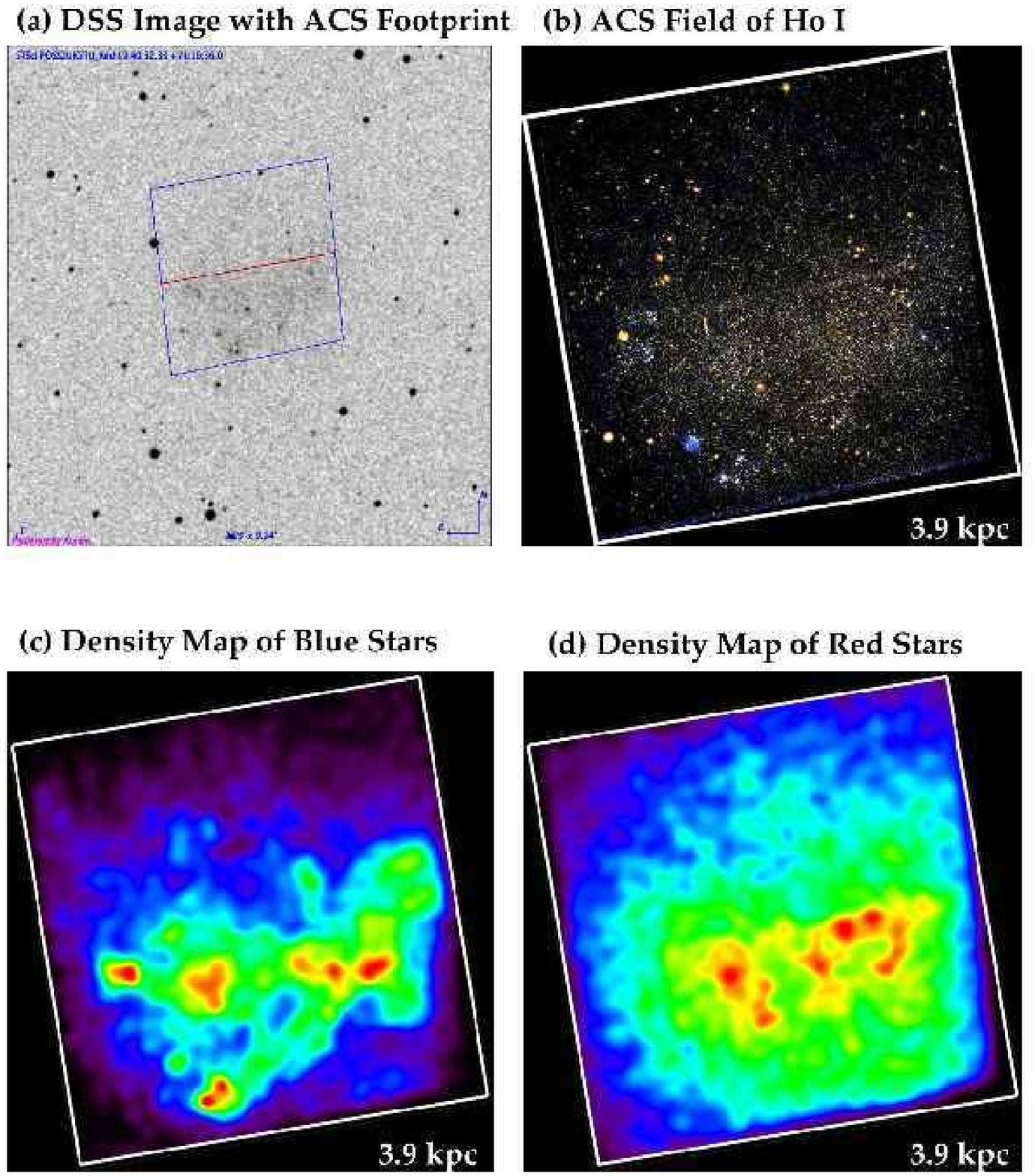}
\caption{A four panel figure of the \hoi\ showing the ACS footprint in blue with the chip gap direction in red over the R-Band DSS image (a), the color ACS image (b), and stellar density maps of blue (c) and red stars (d) as determined by their positions on the CMD.  The blue stars have a color $<$ 0.6 and magnitude $>$ 60\% completeness in F814W.  The red stars are defined as 0.6 $\leq$ color $\leq$ 3.0 and a magnitude range between the TRGB and 60\% completeness in F814W.  The highest density regions are the red contours and the lowest are in purple.}
\label{hoi_image}
\end{center}
\end{figure}

\begin{figure}[t]
\begin{center}
\epsscale{0.9}
\plotone{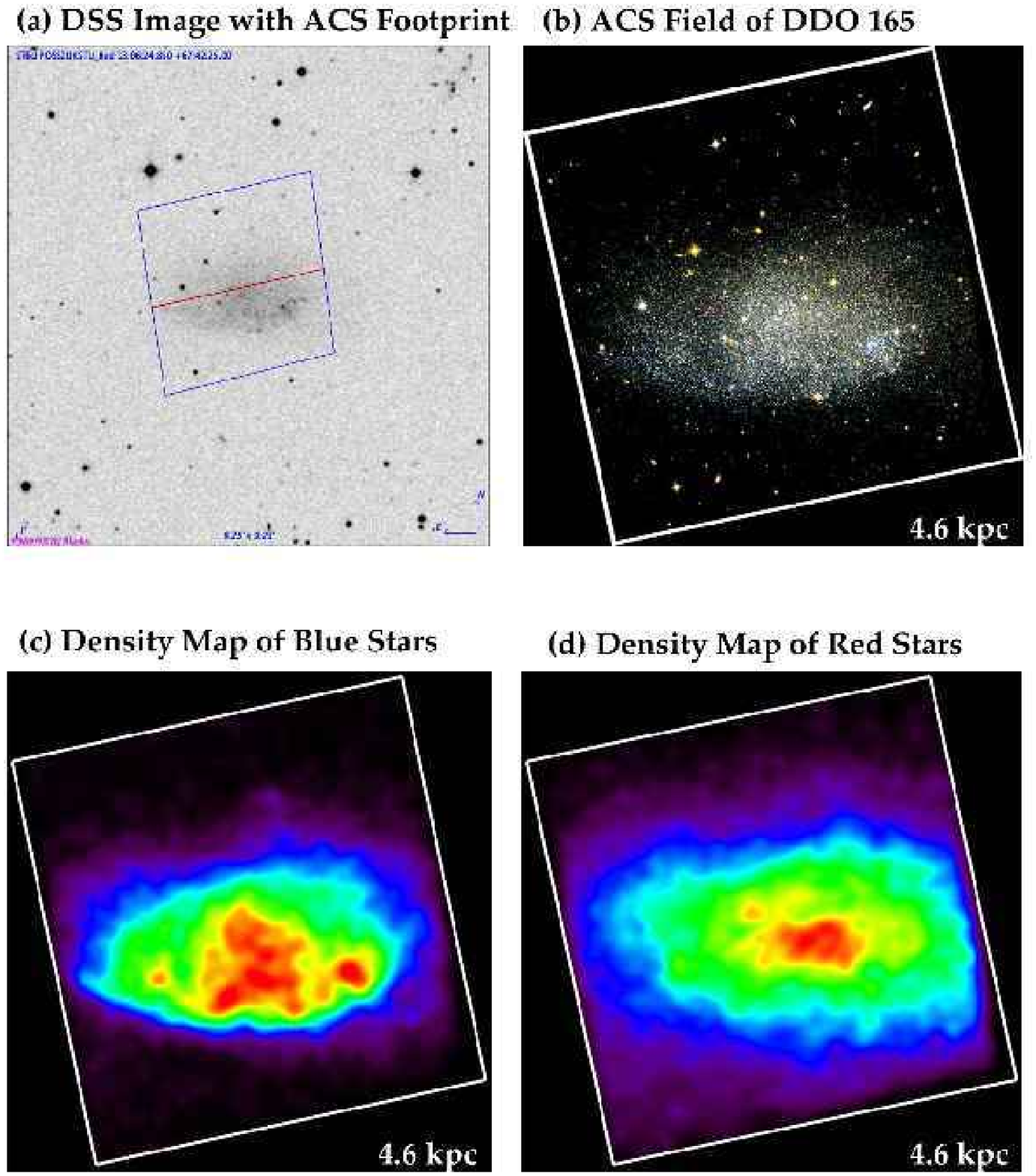}
\caption{A four panel figure of the DDO 165 showing the ACS footprint in blue with the chip gap direction in red over the R-Band DSS image (a), the color ACS image (b), and stellar density maps of blue (c) and red stars (d) as determined by their positions on the CMD.  The blue stars have a color $<$ 0.6 and magnitude $>$ 60\% completeness in F814W.  The red stars are defined as 0.6 $\leq$ color $\leq$ 3.0 and a magnitude range between the TRGB and 60\% completeness in F814W.  The highest density regions are the red contours and the lowest are in purple.}
\label{ddo165_image}
\end{center}
\end{figure}

\begin{figure}[t]
\begin{center}
\epsscale{0.9}
\plotone{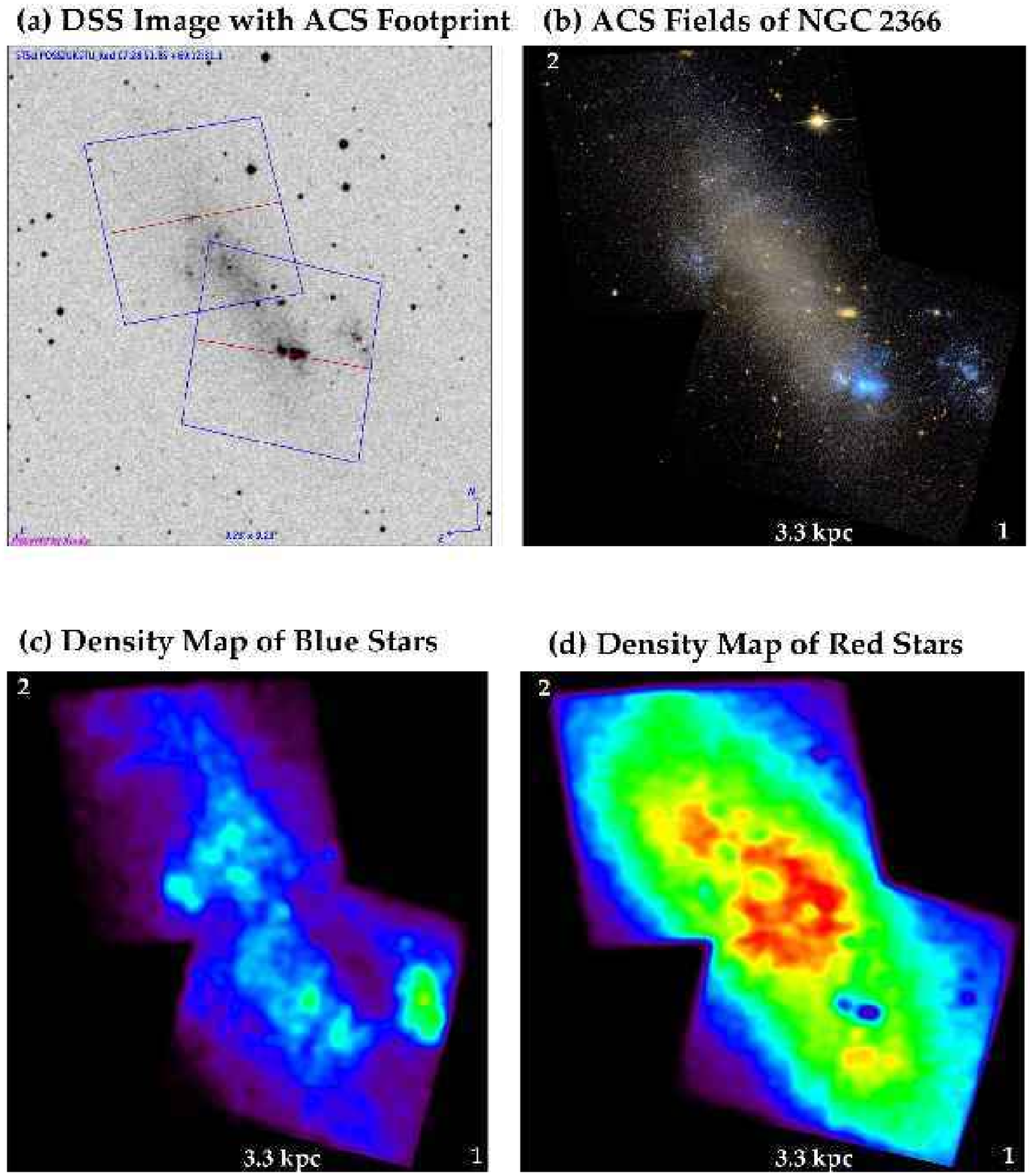}
\caption{A four panel figure of the NGC 2366 showing the ACS footprints in blue with the chip gap directions in red over the R-Band DSS image (a), the color ACS image (b), and stellar density maps of blue (c) and red stars (d) as determined by their positions on the CMD.  The blue stars have a color $<$ 0.6 and magnitude $>$ 60\% completeness in F814W.  The red stars are defined as 0.6 $\leq$ color $\leq$ 3.0 and a magnitude range between the TRGB and 60\% completeness in F814W.  The highest density regions are the red contours and the lowest are in purple.}
\label{ngc2366_image}
\end{center}
\end{figure}

\begin{figure}[t]
\begin{center}
\epsscale{0.9}
\plotone{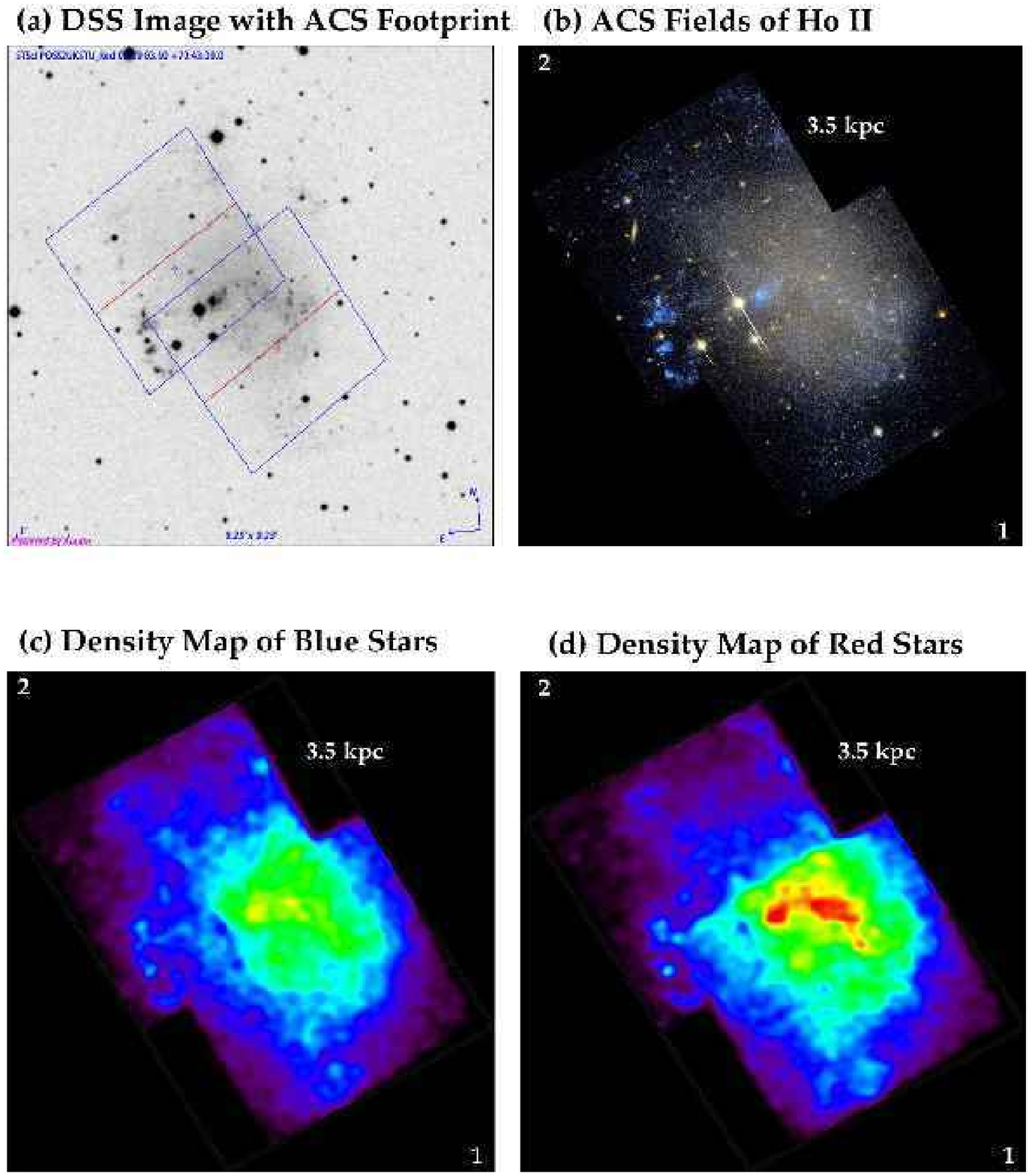}
\caption{A four panel figure of the \hoii\ showing the ACS footprints in blue with the chip gap directions in red over the R-Band DSS image (a), the color ACS image (b), and stellar density maps of blue (c) and red stars (d) as determined by their positions on the CMD.  The blue stars have a color $<$ 0.6 and magnitude $>$ 60\% completeness in F814W.  The red stars are defined as 0.6 $\leq$ color $\leq$ 3.0 and a magnitude range between the TRGB and 60\% completeness in F814W.  The highest density regions are the red contours and the lowest are in purple.}
\label{hoii_image}
\end{center}
\end{figure}

\begin{figure}[t]
\begin{center}
\epsscale{0.9}
\plotone{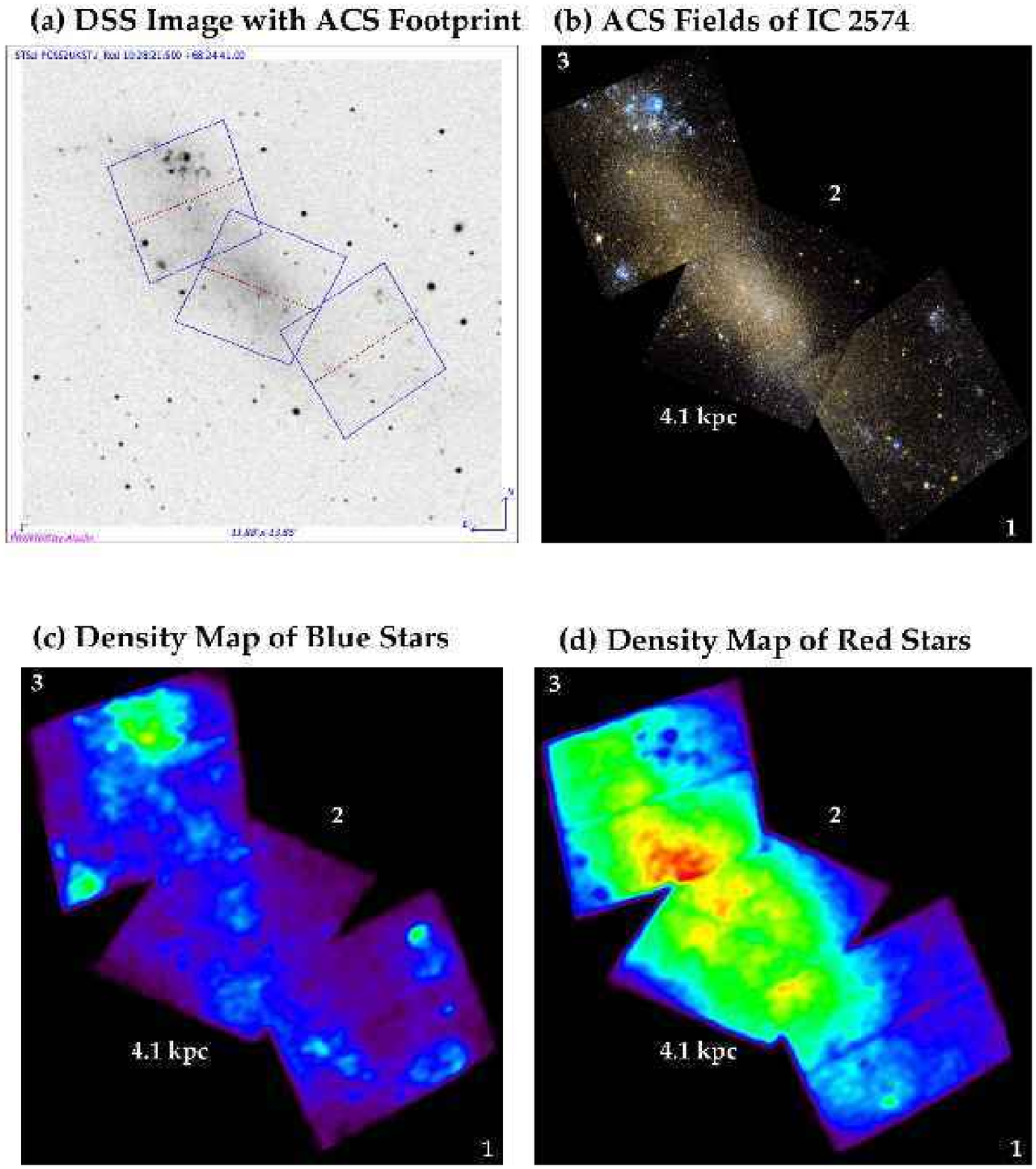}
\caption{A four panel figure of the IC 2574 showing the ACS footprints in blue with the chip gap directions in red over the R-Band DSS image (a), the color ACS image (b), and stellar density maps of blue (c) and red stars (d) as determined by their positions on the CMD.  The blue stars have a color $<$ 0.6 and magnitude $>$ 60\% completeness in F814W.  The red stars are defined as 0.6 $\leq$ color $\leq$ 3.0 and a magnitude range between the TRGB and 60\% completeness in F814W.  The highest density regions are the red contours and the lowest are in purple.}
\label{ic2574_image}
\end{center}
\end{figure}

We present four-panel figures in order to demonstrate provide spatial context of our analysis of each galaxy in our sample (Figures \ref{garland_image} $-$ \ref{ic2574_image}). Panel (a) shows an R-band DSS image with the ACS field(s) of view overlaid in blue with the chip gap position shown in red. Panel (b) shows a color image of each ACS field, created (and combined, for multiple pointings) using drizzled images from the HST pipeline.  Panels (c) and (d) are density maps of the blue (young) and red (old) stars, respectively, in each galaxy.  The density plots of blue and red stars give a sense of the spatial distribution of young and old stars in the each galaxy. Red stars are defined as having a color (i.e., m$_{F555W} $-$ m_{F814W}$) between 0.6 and 3.0 and a magnitude fainter than the TRGB, to exclude the young RHeBs, but include RGB, RC, and AGB stars.  We defined the blue stars to have a color less than 0.6 and include MS and BHeBs.  We also exclude faint spurious detections by considering stars that are only brighter than the 60\% F814W completeness limit as determined by the artificial star tests.  To create the maps, we binned the stars and smoothed the resulting distribution using a Gaussian kernel such that the final image resolution is 7.5\arcsec.

\begin{figure}[t]
\begin{center}
\plotone{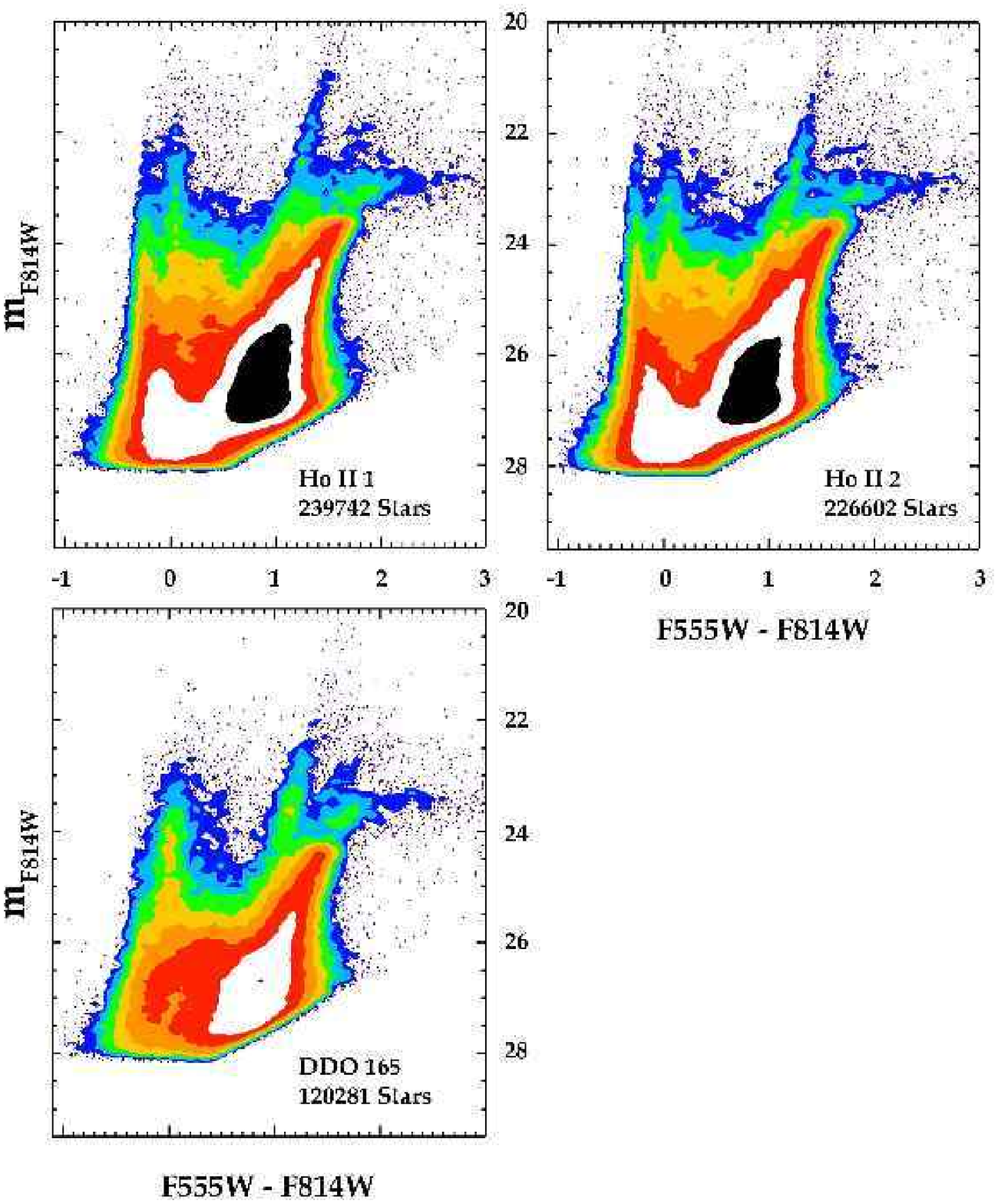}
\caption{The HST/ACS CMDs of both fields of \hoii\  and DDO 165 presented in ACS instrumental filters F555W and F814W, contoured by stellar density on the CMD.  Regions of density less than 4 stars dmag$^{-2}$ (0.1 $\times$ 0.1 magnitudes bins) are plotted as points.  Contours are spaced by factors of 2 to show the detailed structure of the CMDs.}
\label{3cmd1}
\end{center}
\end{figure}

\begin{figure}[t]
\begin{center}
\plotone{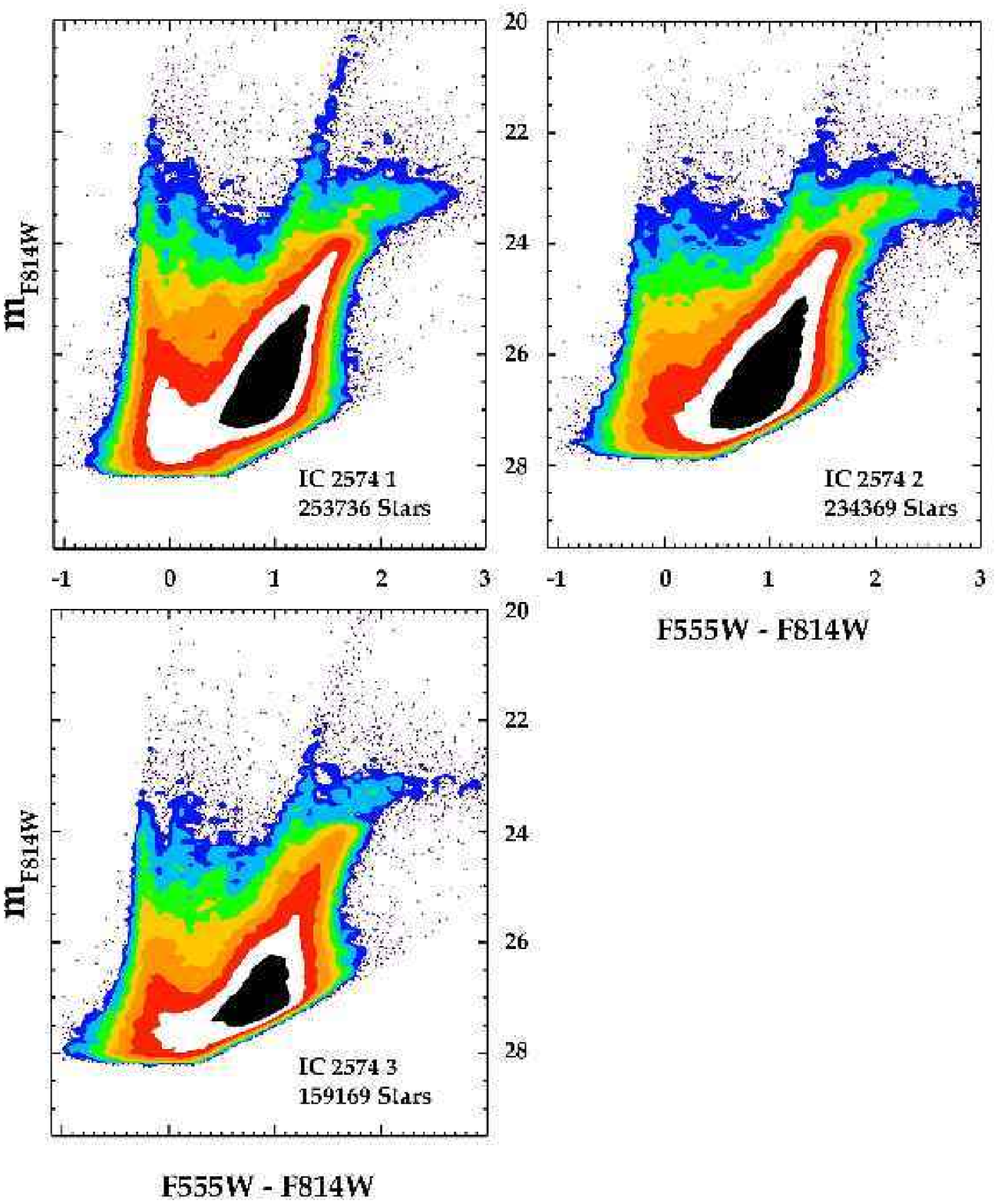}
\caption{The HST/ACS CMDs of all three field of IC 2574 presented in ACS instrumental filters F555W and F814W, contoured by stellar density on the CMD.  Regions of density less than 4 stars dmag$^{-2}$ (0.1 $\times$ 0.1 magnitudes bins) are plotted as points.  Contours are spaced by factors of 2 to show the detailed structure of the CMDs.}
\label{3cmd3}
\end{center}
\end{figure}

The CMDs of each ACS field are shown as density contours to reveal greater detail in otherwise crowded regions of the CMD (Figures \ref{4cmd1} $-$ \ref{3cmd3}).  These CMDs were made using a nearest neighbor plotting algorithm such that regions with a stellar density $\geq$ 4 stars dmag$^{-2}$ (decimags, 0.1 magnitudes) are shown as contours with the contour levels spaced uniformly by a factor of 2.  Stars are plotted as individual points when the density $<$ 4 stars dmag$^{-2}$.

The lifetime (Figures \ref{global_small} and \ref{global_big}) and recent (Figures \ref{recent_small} and \ref{recent_big}) SFHs show the SFHs of each galaxy for both the history of the universe and the most recent 1 Gyr with different time resolutions. The typical time resolution for the lifetime SFHs is $\Delta$ $\log(t)$ $\sim$ $0.3$, while the recent SFHs have a 10 Myr resolution in the most recent time bin and a $\sim$ 250 Myr resolution for the oldest time bin. We note that the recent SFHs appear different from the most recent 1 Gyr of the lifetime SFHs due to the choice of binning, although they are equivalent SFHs.    We chose a finer time resolution for the most recent 1 Gyr because of excellent leverage on the recent SF afforded by the BHeBs, allowing us to study pattens of recent SF in greater detail.  

Similar to the different appearance of the lifetime and recent SFHs, the value of the SFR error bars also vary with time bin resolution.  This effect is most evident in the oldest age bin of the lifetime SFHs.  Although the photometry only allows us to place weak constraints on the ancient SFHs, the error bar appears quite small.  This is because we bin all SF $>$ 6 Gyr into a single coarse bin as opposed to multiple bins of finer time resolution.  If we had selected finer time resolution, we would have increased the uncertainty in the SFR in each time bin.  Despite being able to quantify the ancient SFH, we can only do so with very low time resolution (6 $-$ 14 Gyr), which equates to placing a loose constraint on SF at this epoch.

To quantitatively compare the SFHs, we employ the two parameters.  The first, $f$, is the fraction of stars formed over a given time period.  Specifically, we compute $f_{1}$, $f_{6}$, and $f_{14}$, the fraction of stars formed from 0 $-$ 1 Gyr, 1 $-$ 6 Gyr, and 6 $-$14 Gyr (see Table \ref{tab3}).  By comparing the fraction of stars we can assess how active the galaxy has been at recent, intermediate, and ancient times (e.g., for a constant SFR, $f_{1}$ $=$ 0.07, $f_{6}$ $=$ 0.36, and $f_{14}$ $=$ 0.57).  The second set of parameters we compute are the  `birthrate' parameters,
i.e., the ratio of the current SFR to the lifetime averaged SFR
\citep[cf.,][]{Sca86,ken98}, for recent time intervals of 100 Myr, 500 Myr and 1 Gyr.
The birthrate parameter $b_{100}$ represents the SFH of a galaxy averaged over the past 100 Myr normalized to the lifetime average, and, similarly, $b_{500}$ and $b_{1G}$ are the birthrate parameters for the past 500 Myr and 1 Gyr respectively (see Table \ref{tab3}).  A galaxy that forms stars at a constant rate would have a birthrate parameter of unity.  Birthrate parameters greater than unity show  that the SFR is a higher than average at that time.

\subsection{Galaxies with M$_{B}$ $>$ $-$15}

 In this section we present the CMDs, spatial stellar distribution, and SFHs for each of the galaxies with M$_{B}$ $>$ $-$15 in increasing order of luminosity.  We discuss and compare the formation and evolution of these galaxies in \S 6.

\begin{figure}[t]
\begin{center}
\epsscale{0.85}
\plotone{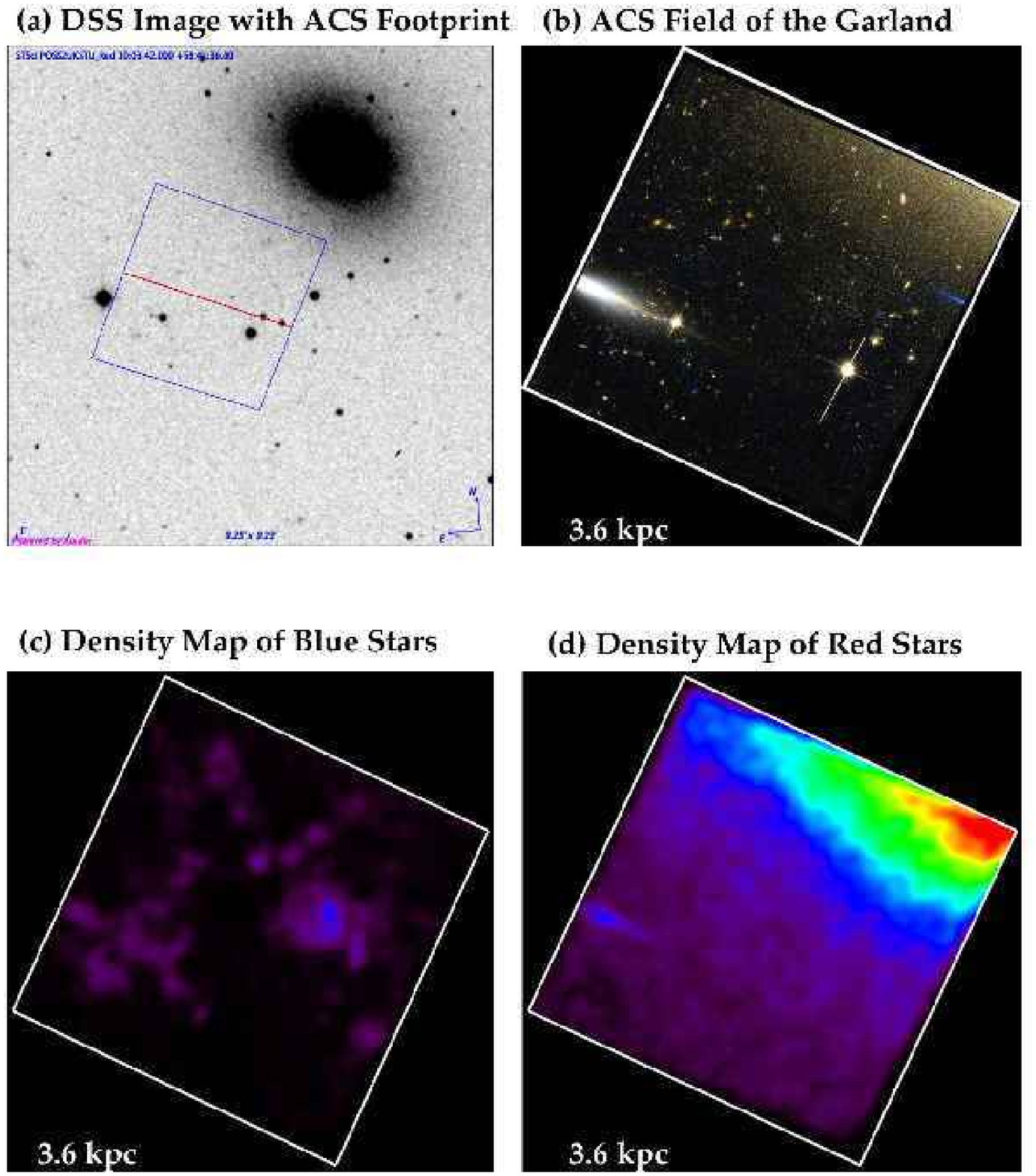}
\caption{A four panel figure of the Garland showing the ACS footprint in blue with the chip gap direction in red over the R-Band DSS image (a), the color ACS image (b), and stellar density maps of blue (c) and red stars (d) as determined by their positions on the CMD.  The blue stars have a color $<$ 0.6 and magnitude $>$ 60\% completeness in F814W.  The red stars are defined as 0.6 $\leq$ color $\leq$ 3.0 and a magnitude range between the TRGB and 60\% completeness in F814W.  The highest density regions are the red contours and the lowest are in purple.  Note our ACS observation of the Garland contains a significant number of red stars from NGC 3077 (see \S 4.1 for further discussion).}
\label{garland_image}
\end{center}
\end{figure}

\textbf{The Garland --}  As shown in Figure \ref{garland_image}, the Garland is located in close proximity to the southeast of NGC 3077.  The Garland is very hard to discern in both the DSS and color ACS images, but is more prevalent in the density maps.  There is a dense red star component (panel (d)) in the ACS field which decreases away from NGC 3077 implying that these stars are associated with the larger galaxy.  In contrast, the blue stellar density map (panel (c)) reveals several regions of concentrated blue stars in agreement with \citet{sak01}.  The region of highest blue star concentration overlaps with the the outer red stars and it is not possible from the ACS observations alone to determine if these blue stars are a separate system or merely young stars in the outer part of NGC 3077.  We also note that there does not appear to be an overdensity of red stars associated with any of the concentrations of blue stars.

\begin{figure}[t]
\begin{center}
\plotone{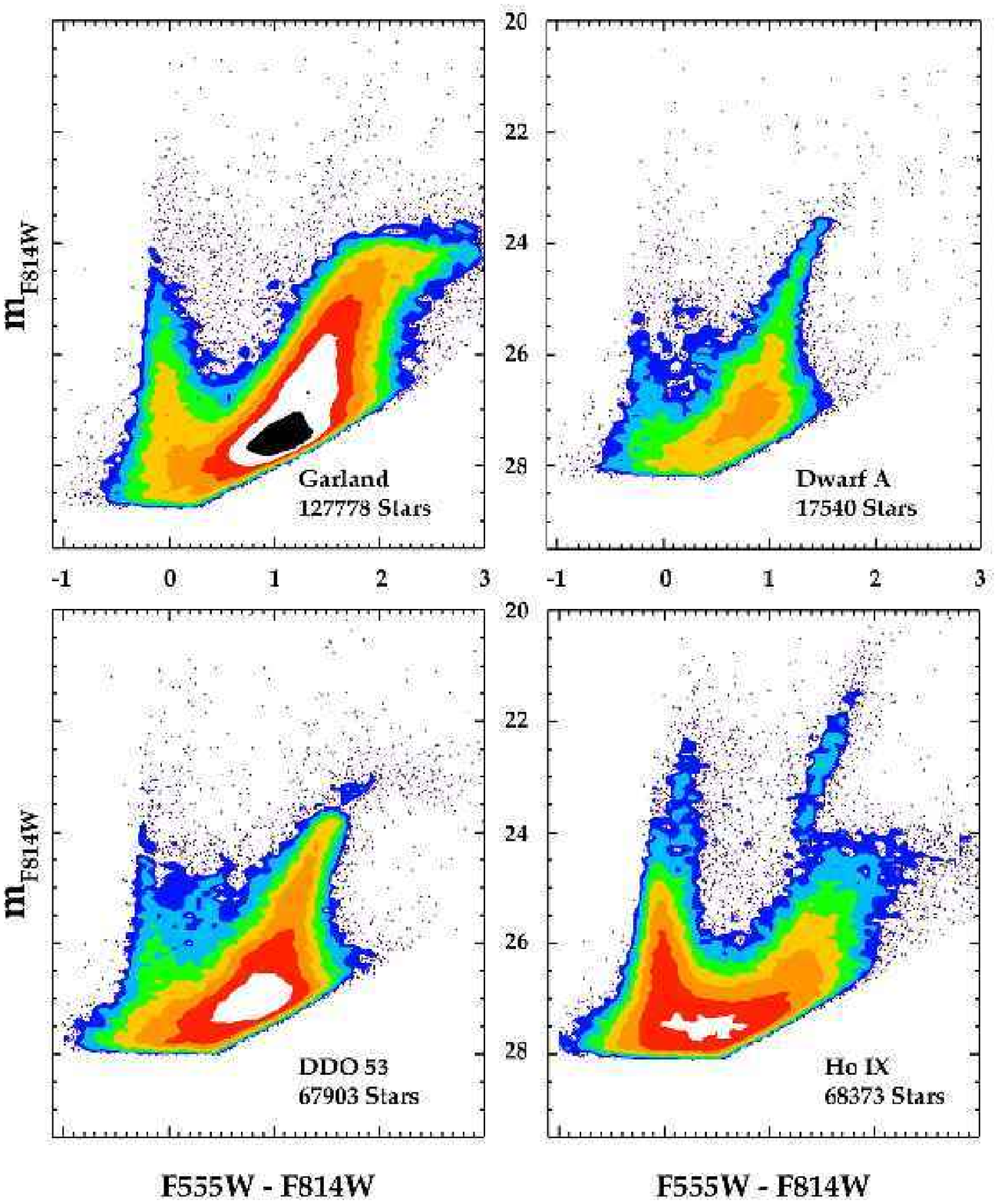}
\caption{The HST/ACS  CMDs of M81 Dwarf~A, the Garland, DDO 53, and \hoix\ presented in ACS instrumental filters F555W and F814W, contoured by stellar density on the CMD.  Regions of density less than 4 stars dmag$^{-2}$ (0.1 $\times$ 0.1 magnitudes bins) are plotted as points.  Contours are spaced by factors of 2 to show the detailed structure of the CMDs.  }
\label{4cmd1}
\end{center}
\end{figure}

In Figure \ref{4cmd1} we show the CMD of the entire ACS field of the Garland.  The high stellar density and breadth of the RGB are due to the red stars from the outer part of NGC 3077.  Young stars are present in the form of MS stars and a very sparse distribution of HeBs. 

\begin{sidewaysfigure}[t]
\begin{center}
\includegraphics[scale=0.8, angle=270]{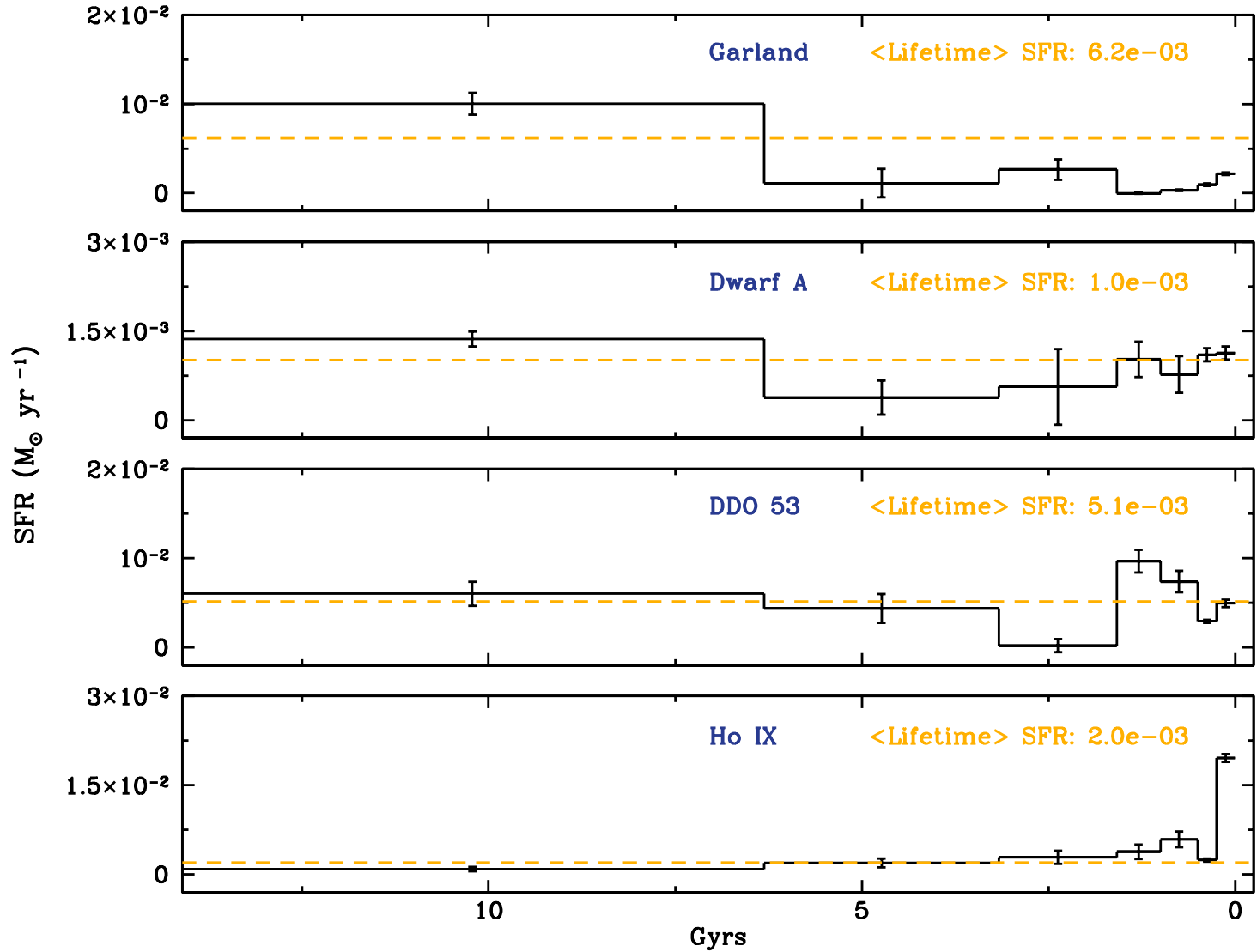}
\caption{The lifetime SFHs of the Garland, M81 Dwarf~A, \hoix, and DDO 53 with a time resolution of $\sim$ $\log(t)$ $\sim$ 0.3.  The orange dashed line represents the SFH averaged over the lifetime of the of galaxy.  The error bars reflect both systematic and statistical uncertainties.}
\label{global_small}
\end{center}
\end{sidewaysfigure}

\begin{sidewaysfigure}[t]
\begin{center}
\includegraphics[scale=0.8, angle=270]{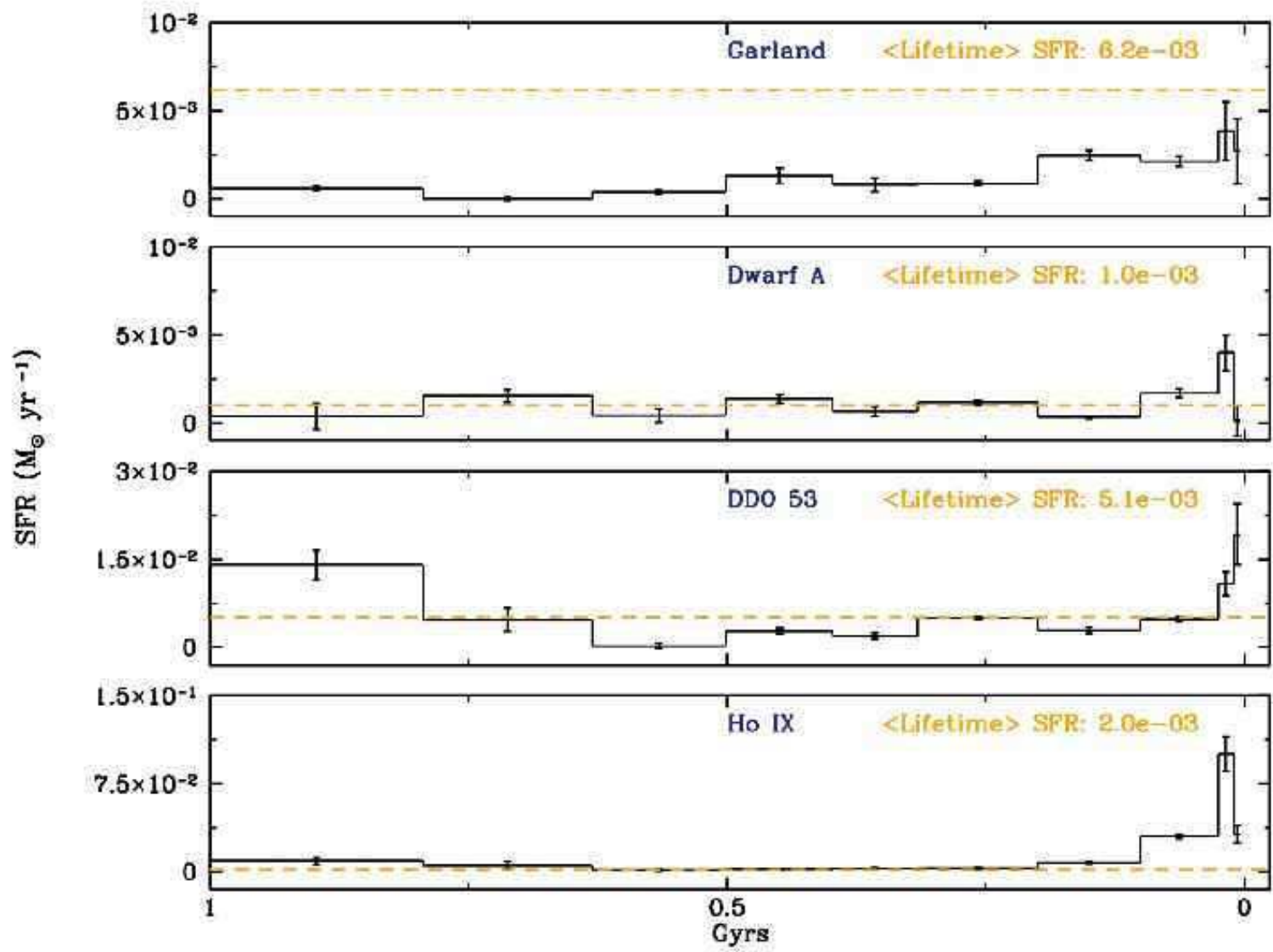}
\caption{The recent SFHs of the Garland, M81 Dwarf~A, \hoix, and DDO 53 with a time resolution ranging from 10 Myr in the most recent time bin to $\sim$ 250 Myr in the oldest bin.  The orange dashed line represents the SFH averaged over the lifetime of the of galaxy.  The error bars reflect both systematic and statistical uncertainties.}  \label{recent_small}
\end{center}
\end{sidewaysfigure}

The lifetime and recent SFHs of the Garland are shown in Figures \ref{global_small} and \ref{recent_small}. To produce a more accurate SFH, we isolated the Garland from NGC 3077 by subtracting out the stars that lie in the green through red contour shown in panel (d) of Figure \ref{garland_image}. 

The lifetime SFH shows an ancient SFR which exceeds the average lifetime SFR ($\sim$ 0.006 \msun\ yr$^{-1}$) and is presumably due to the inclusion of red stars from NGC 3077; only 1\% of the stars in the ACS image have been formed within the more recent Gyr.  The additional red stars have the effect of increasing the lifetime average SFR of the Garland.  Over the most recent 1 Gyr, we see that the SF remains roughly constant until $\sim$ 300 Myr ago when there was an increase in the SFR that continues to the present.  We quantify this  increase in SFR by computing  $\frac{b_{100}}{b_{1G}}$ $\sim$ 2.5, done so to eliminate the dependence on the contaminated lifetime average SFR (see Table \ref{tab3}). To further aid in measuring the SFH, we did not require the metallicity to monotonically increase with time.  The Garland is best fit with more metal poor isochrones over the past 1 Gyr (\mh\ $=$ $-1.0$) and more metal rich ancient (\mh\ $=$ $-0.4$) and intermediate age (\mh\ $=$ $-0.3$) components.  This metallicity distribution is consistent with the old and intermediate age stars coming from the more massive NGC 3077, while the young stars in the Garland come from metal poor gas.

\begin{figure}[t]
\begin{center}
\epsscale{0.9}
\plotone{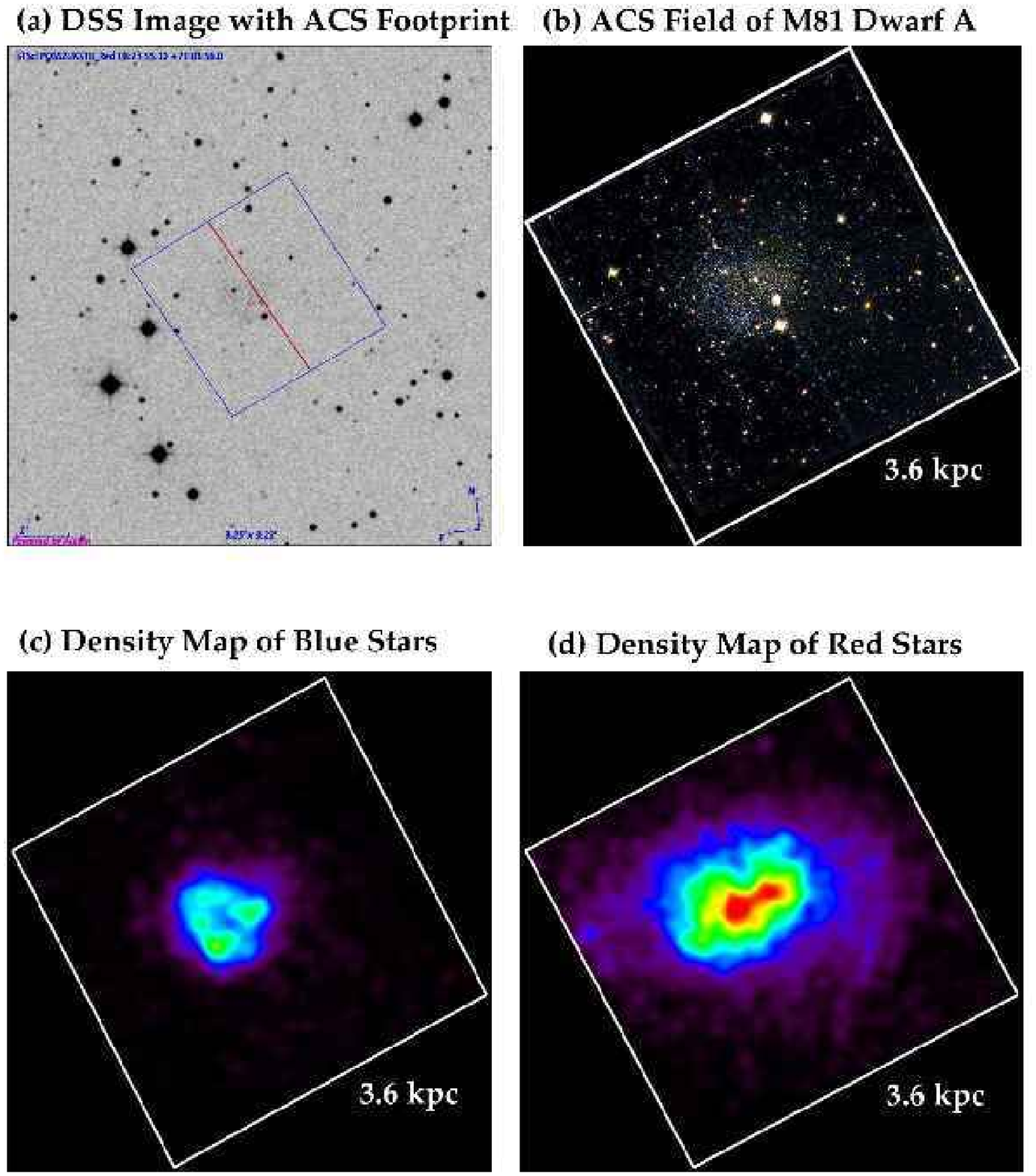}
\caption{A four panel figure of the M81 Dwarf~A showing the ACS footprint in blue with the chip gap direction in red over the R-Band DSS image (a), the color ACS image (b), and stellar density maps of blue (c) and red stars (d) as determined by their positions on the CMD.  The blue stars have a color $<$ 0.6 and magnitude $>$ 60\% completeness in F814W.  The red stars are defined as 0.6 $\leq$ color $\leq$ 3.0 and a magnitude range between the TRGB and 60\% completeness in F814W.  The highest density regions are the red contours and the lowest are in purple.}
\label{dwarfa_image}
\end{center}
\end{figure}

\textbf{M81 Dwarf~A --} The ACS observation and stellar density maps of M81 Dwarf~A are shown in Figure \ref{dwarfa_image}.   We see from panel (d) in Figure \ref{dwarfa_image} that the red stars are highly concentrated in the center, with a slight elongation in the southeast-northwest direction.  The highest blue star concentrations form a ring shape and are aligned with the densest region of rest stars (panel (c)). 

From the CMD of M81 Dwarf~A (Figure \ref{4cmd1}) we see a prominent RGB and very few AGB stars.  The MS and BHeB sequences are distinct and both moderately populated, although their absence at brighter magnitudes suggests that there is little, if any, recent SF.

The lifetime and recent SFHs of M81 Dwarf~A are shown in Figures \ref{global_small} and \ref{recent_small}.  The lifetime SFH shows the oldest stars were formed at a rate higher than the lifetime average (0.001 \msun\ yr$^{-1}$), $f_{14}$ $=$ 0.75, followed by intermediate age SF at a rate lower than the lifetime average ($f_{6}$ $=$ 0.19) until the last $\sim$ 2 Gyr.  Within the last 1 Gyr the SFH is consistent with the lifetime average, $b_{1G}$ $=$ 0.92 and $b_{500}$ $=$ 1.08, with a significant increase in the SFR in the past $\sim$ 100 Myr ($b_{100}$ $=$ 1.84).  In the most recent time bin, 0 $-$ 10 Myr, the SFR significantly drops off, which is confirmed by the absence of \halpha\ emission, indicating M81 Dwarf~A is not forming stars at the present \citep{leethesis, ken08}.

\textbf{DDO 53 --} Figure \ref{ddo53_image} shows the ACS observations and stellar density maps of DDO 53.  The red stellar density map, shown in panel (d), reveals an elongated distribution of red stars in the northeast-southwest direction that appears to extend to the east beyond the ACS field of view.  The highest concentration of blue stars (panel (c)) is coincident with the highest concentration of red stars, and does not seem to suffer from truncation as the red stars do.  The asymmetric distribution of red stars could be due to an interaction and is discussed in greater detail in \S 6.

The CMD of DDO 53 (Figure \ref{4cmd1}) shows high density RC and RGB stars, with a smaller population of AGB stars.  There is a bright MS of moderate stellar density and HeBs are present but more so at fainter magnitudes, which are indicative of recent SF ($<$ 25 Mrs).

The lifetime and recent SFHs are shown in Figures \ref{global_small} and \ref{recent_small}.  The lifetime SFH shows ancient and intermediate SF consistent with the lifetime average of $\sim$ 0.005 \msun\ yr$^{-1}$ ($f_{14}$ $=$ 0.64 and $f_{6}$ $=$ 0.28).  Over the most recent 1 Gyr DDO 53 has formed 8\% of its stars and the birthrate parameters are very near the lifetime average, except over the past 100 Myr ($b_{100}$ $=$ 1.24, see Table \ref{tab3}).  The cause of this rise in birthrate is due to an elevated SFR within the past 25 Myr, which is in excellent agreement with the presence of more young MS than HeB stars, the latter of which do not appear before 25 Myr.

\begin{figure}[t]
\begin{center}
\epsscale{0.85}
\plotone{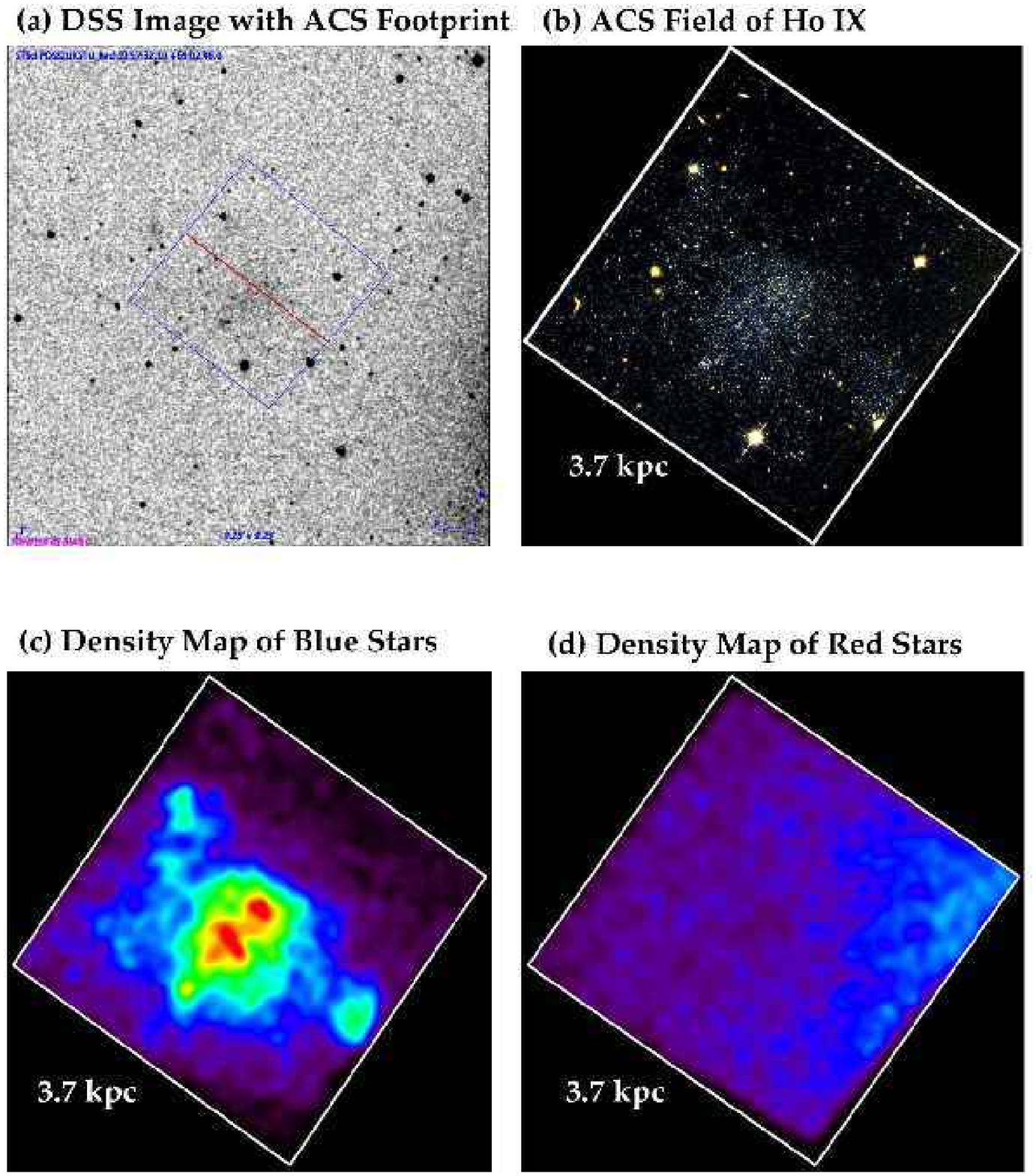}
\caption{A four panel figure of the \hoix\ showing the ACS footprint in blue with the chip gap direction in red over the R-Band DSS image (a), the color ACS image (b), and stellar density maps of blue (c) and red stars (d) as determined by their positions on the CMD.  The blue stars have a color $<$ 0.6 and magnitude $>$ 60\% completeness in F814W.  The red stars are defined as 0.6 $\leq$ color $\leq$ 3.0 and a magnitude range between the TRGB and 60\% completeness in F814W.  The highest density regions are the red contours and the lowest are in purple.  Note that our ACS field of \hoix\ contains red stars from M81 (see \S 4.1 for further discussion).}
\label{hoix_image}
\end{center}
\end{figure}

\textbf{Holmberg IX --}  The blue and red stellar density maps reveal a wealth of information about \hoix\ (see Figure \ref{hoix_image}).  The highest concentration of red stars (panel (d)) is along the western edge of the ACS field and decreases in density across the field to the east.  It is unclear whether the low level of red stars (purple contours in panel (d)) that occupy the majority of the field are due to background or the very outer reaches of M81.  The blue stars (panel (c)) are very prominent in the blue stellar density map as they show a very high density concentration of stars in the center of the ACS field.   Similar to the Garland, we do not find a concentration of red stars associated with the blue stars of \hoix.

In Figure \ref{4cmd1} we show the CMD of the entire ACS field for \hoix.  Much like the Garland, \hoix\ has a relatively broad RGB, which is due to contamination from red stars associated with M81.  What makes the CMD of \hoix\ especially interesting is the clustered dense nature of the HeBs, which is the signature of a recent burst of SF activity.  The lack of bright MS stars tells us that \hoix\ very recently stopped forming stars.

The lifetime and recent SFHs of \hoix\ are shown in Figures \ref{global_small} and \ref{recent_small}.  In a similar manner as for the Garland, we excluded the stars associated with M81 (roughly, we removed most of the blue and cyan contoured stars in panel (d) of Figure \ref{hoix_image}) and measured the SFH of \hoix\ based on the remaining stars.  We also did not require a monotonically increasing metallicity in the CMD fitting routine.  Like the Garland, \hoix\ is best fit with a more metal rich old component and a more metal poor young component (see Table \ref{tab3}).

The lifetime SFH shows that \hoix\ has ancient and intermediate age SFRs consistent with the lifetime average SFR ($\sim$ 0.002 \msun\ yr$^{-1}$).  We find that \hoix\ formed a significant fraction of its stars within the past 1 Gyr ($f_{1}$ $=$ 0.30) with SFRs up to the most recent $\sim$ 250 Myr consistent with the lifetime average SFR.  Even more recently, we see that $b_{100}$ $\sim$ 20, which implies a large fraction of stellar mass in \hoix\ was formed within the past 100 Myr.  The most recent time bin, 0 $-$ 10 Myr, shows a significant decline in the SFR, and \halpha\ observations show an anomalously low \halpha\ EW of 9\AA , which is the signature of a post-burst galaxy  \citep{leethesis, ken08}.   The combined SFH and \halpha, which is sensitive to $\sim$ 0 $-$ 6 Myr SF, are evidence that \hoix\ is in a post-burst state and has just stopped forming stars within the last $\sim$ 10 Myr.  We discuss the formation and evolution of \hoix\ in more detail in \S 6.


\textbf{\hoi\  --}  Figure \ref{hoi_image} shows the ACS observation and stellar density maps of \hoi.  The ACS field has captured most of the galaxy, although the red stellar distribution appears to be truncated to the south and west (panel (d)).  There is a high density concentration of red stars in the center of the field, with a more moderate density component occupying the rest of the field.  The blue stars are centrally located in the field, with several areas of very high density (panel (c)).  

The CMD of \hoi\ is shown in Figure \ref{3cmd2}.  The RGB and RC are densely populated due to ancient SF and the presence of a more sparsely populated AGB sequence implies lower levels of intermediate age SF.  The HeB and MS tracks share roughly the same densities and together tell us the \hoi\ has recently been forming stars.

The lifetime and recent SFHs of \hoi\ are shown in Figures \ref{global_big} and \ref{recent_big}. The ancient SFR is slightly greater than the lifetime average SFR ($\sim$ 0.01 \msun\ yr$^{-1}$), while the intermediate age SFRs are lower than the lifetime average SFR ($f_{14}$ $=$ 0.76, $f_{6}$ $=$ 0.15).  Within the past 1 Gyr, we find the SFRs are slightly greater ($b_{1G}$ $=$ 1.26) than the lifetime average and that the most significant deviation from the lifetime average SFR occurs within the past 100 Myr, $b_{100}$ $=$ 1.88.  \halpha\ observations confirm that \hoi\ is presently forming stars \citep{ott01, leethesis, ken08}.

\subsection{Galaxies with M$_{B}$ $\leq$ $-$15}

 In this section we present the CMDs, spatial stellar distribution, and SFHs for each of the galaxies with M$_{B}$ $\leq$ $-$15in order of increasing luminosity.  We do not provide extensive comments and comparison with other wavelengths on individual features in the larger galaxies. In future work, we will conduct more detailed analysises of the spatial resolved recent SFHs of these galaxies.

\textbf{DDO 165 --}  Figure \ref{ddo165_image} shows the ACS observation and stellar density maps of DDO 165.  The red stellar density map (panel (d)) shows a distribution of red stars elongated in the east-west direction, with the highest density centrally located.  The red stars appear to extend beyond the ACS field of view on the west side.  There is a high density central concentration of blue stars (panel (c)) that cover a larger area than the central concentration of red stars.  The blue stars are also elongated along the east-west axis.

The CMD of DDO 165 is shown in Figure \ref{3cmd1}.  We see fairly well populated RGB and AGB tracks, indicators of old and intermediate SF.  The young stars in this galaxy are particularly interesting and unusual.  On the blue side, we do not see separate MS and BHeB tracks, rather, there is just one fairly dense blue star feature.  We are able to conclude that this feature is indeed composed of BHeBs based on the empirical work of \citet{dom02}, which show that in Local Group dI Sextants A, that the ratio of BHeBs to RHeBs is greater than unity on average.  The fact that the HeBs are so dominant over the MS is evidence that DDO 165 must have had recent SF activity, but is no longer forming stars.

\begin{sidewaysfigure}[t]
\begin{center}
\includegraphics[scale=0.8, angle=270]{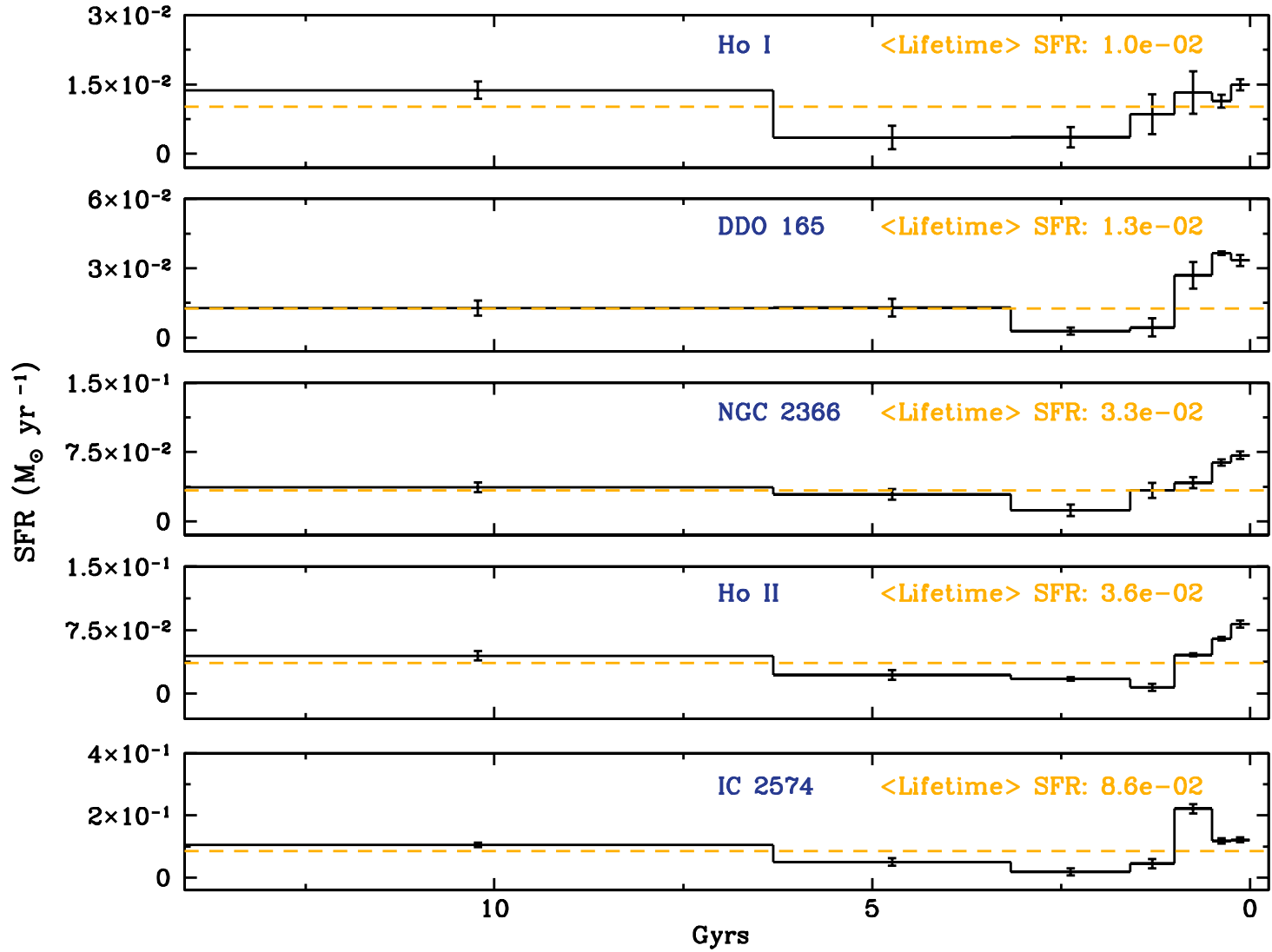}
\caption{The lifetime SFHs of \hoi, DDO 165, NGC 2366, \hoii, and IC 2574 with a time resolution of $\sim$ $\log(t)$ $\sim$ 0.3.  The orange dashed line represents the SFH averaged over the lifetime of the of galaxy.  The error bars reflect both systematic and statistical uncertainties.}  \label{global_big}
\end{center}
\end{sidewaysfigure}

\begin{sidewaysfigure}[t]
\begin{center}
\includegraphics[scale=0.8, angle=270]{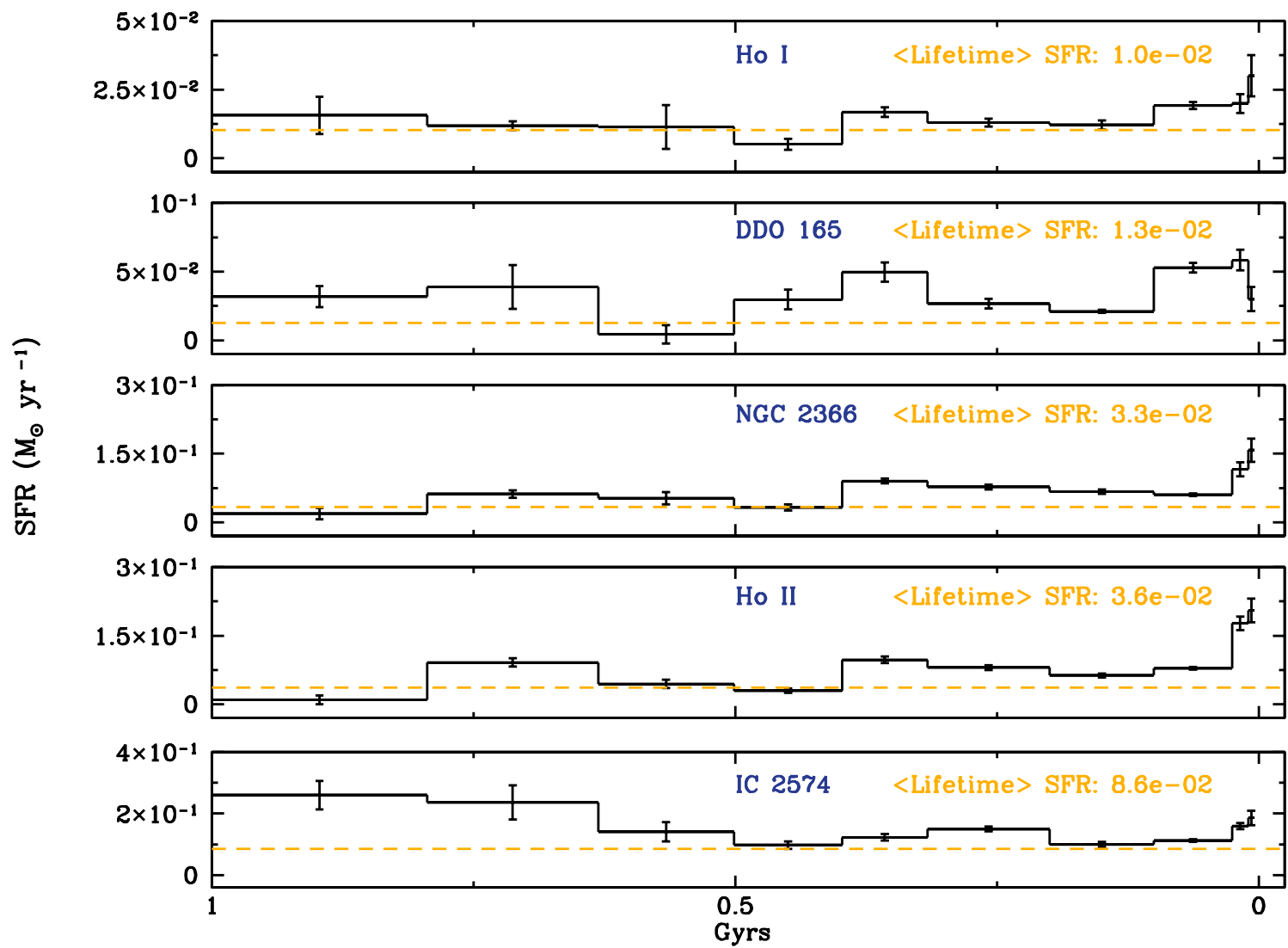}
\caption{The recent SFHs of \hoi, DDO 165, NGC 2366, \hoii, and IC 2574 with a time resolution ranging from 10 Myr in the most recent time bin to $\sim$ 250 Myr in the oldest bin.  The orange dashed line represents the SFH averaged over the lifetime of the of galaxy.  The error bars reflect both systematic and statistical uncertainties.}  
\label{recent_big}
\end{center}
\end{sidewaysfigure}

The lifetime and recent SFHs of DDO 165 are shown in Figures \ref{global_big} and \ref{recent_big}.  The ancient and intermediate age SFRs are mostly consistent with the lifetime average SFR ($\sim$ 0.013 \msun\ yr$^{-1}$) with $f_{14}$ $=$ 0.55 and $f_{6}$ $=$ 0.28.  Over the past 1 Gyr, DDO 165 has formed 17\% of its stars, the highest fraction in our sample outside of \hoix\ (Table \ref{tab3}).   We see that the birthrate parameters are consistently higher the unity ($b_{1G}$ $=$ 2.43, $b_{500}$ $=$ 2.74, and $b_{100}$ $=$ 3.95), with a significant increase over the past 100 Myr.   There is little \halpha\ emission in DDO 165, indicating very little, if any current SF.  Further, the low \halpha\ equivalent width (EW) classifies DDO 165 as post-burst, in excellent agreement with our SFH \citep{leethesis, ken08}.

\textbf{NGC 2366 --}  Two ACS fields were required to cover most of NGC 2366 (Figure \ref{ngc2366_image}).  We created a mosaic of the entire galaxy from the ACS observations (panel (b)) using Multidrizzle \citep{koe02} .  The red stellar density map (panel (d)) reveals an elliptical distribution of red stars with the major axis in the northeast-southwest direction with a central high density concentration.  The blue stars (panel (c)) are scattered through the main body, with a few highly concentrated regions. Field 1 (southern field) contains the powerful star forming \hii\ region, NGC 2363 \citep{kbh80}, seen as the large blue central region in panel (b). Coincident with this region is a dense concentration of blue stars (panel (c)) and a deficiency of red stars (panel (d)).  This may either be a true lack of red stars or, more likely, a photometric effect, due to the high stellar density and extended \hii\ emission in the F555W filter.  To the west of NGC 2366, we find another concentration of blue stars, with no associated clustering of red stars, which is unlikely due to a photometric effect, because the region of extended \hii\ emission is not very large. 

\begin{figure}[t]
\begin{center}
\plotone{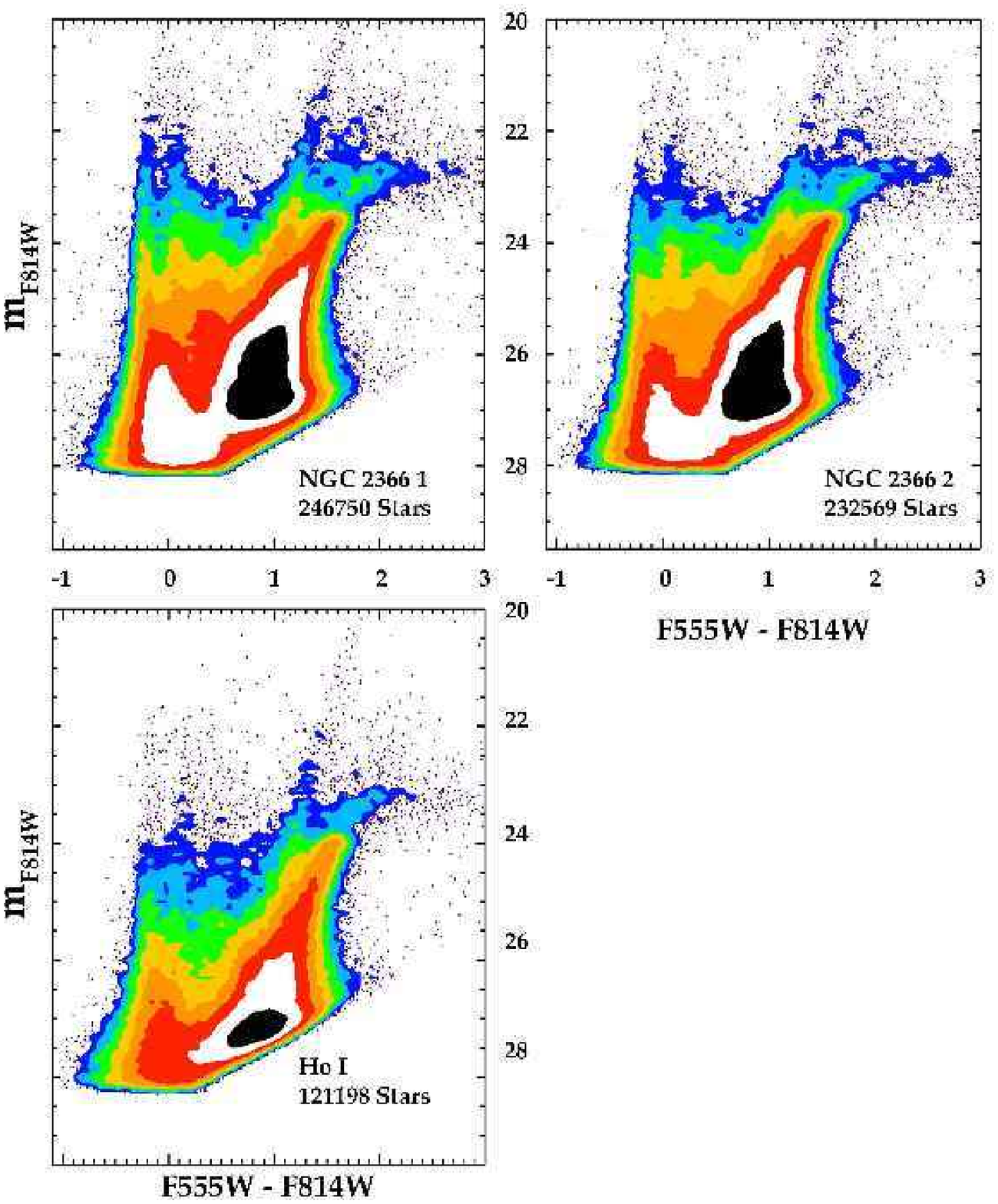}
\caption{The HST/ACS CMDs of both fields of NGC 2366 and \hoi\  presented in ACS instrumental filters F555W and F814W, contoured by stellar density on the CMD.  Regions of density less than 4 stars dmag$^{-2}$ (0.1 $\times$ 0.1 magnitudes bins) are plotted as points.  Contours are spaced by factors of 2 to show the detailed structure of the CMDs.}
\label{3cmd2}
\end{center}
\end{figure}

We show the CMDs of Fields 1 and 2 in Figure \ref{3cmd2}.  The similarity of the RGB, RC, and AGB components of the two fields reflect the symmetry of the spatial red star distribution in the galaxy.  Field 1, which contains NGC 2363, contains MS and HeB tracks that are more densely populated at brighter magnitudes, indicating the southern half of the galaxy has stronger recent SF.  

The lifetime and recent SFHs of NGC 2366 are shown in Figures \ref{global_big} and \ref{recent_big} for Fields 1 and 2 combined.  The ancient and intermediate SFRs are mostly consistent with the lifetime average SFR ($\sim$ 0.03 \msun\ yr$^{-1}$) with $f_{14}$ $=$ 0.59 and $f_{6}$ $=$ 0.29.  Over the past 1 Gyr, the birthrate parameters range from $\sim$ 1.5 $-$ 2 and $f_{1}$ $=$ 0.12, indicating NGC 2366 has been very actively forming stars.

\textbf{\hoii\  --} \hoii\ required two ACS fields to cover a significant fraction of the galaxy (Figure \ref{hoii_image}). The color ACS figure (panel (b)) was created using Multidrizzle.   Even with the two fields, there is still some truncation of both blue (panel (c)) and red (panel (d)) stars in the northwest direction, due to the galaxy extending beyond the ACS fields.  Neither the red nor blue stars show preferential elongation along any axis.  The red stars are centrally concentrated and coincident with the highest concentration of blue stars.  In the color ACS image (panel (b)), we see several prominent blue regions of \hii\ emission on the eastern side of the galaxy.  Each of these blue regions has an associated overdensity of  blue and red stars, suggesting these regions have been forming stars more than just at recent times.

The CMDs of Field 1 (south) and 2 (north) of \hoii\ are shown in Figure \ref{3cmd2}.  The RC, RGB, and AGB sequences are similarly populated, which is in line with the symmetry of the galaxy's red stellar spatial distribution (panel (d)).  The MS and HeBs are also virtually identical on the CMD, although Field 2 contains a more densely populated bright MS, indicative of stronger recent SF in Field 2.

The lifetime and recent SFHs of the combined fields of \hoii\ are shown in Figures \ref{global_big} and \ref{recent_big}.   The ancient and intermediate SFRs of \hoii\ are consistent with or below the lifetime average ($\sim$ 0.04 \msun\ yr$^{-1}$) with $f_{14}$ $=$ 0.67 and $f_{6}$ $=$ 0.21.  The birthrate parameters of \hoii\ range from 1.6 $-$ 2.7 and $f_{1}$ $=$ 0.12 (see Table \ref{tab3}).   We see a dramatic rise in the SFR over the past 50 Myr.

\textbf{IC 2574 --}   IC 2574 is the largest and most metal rich of the dIs in the sample. We used two ACS fields to cover the southern and central regions and added a previous observation from the HST archive to the north to give us close to full coverage (Figure \ref{ic2574_image}).  We combined the three fields using Multidrizzle to create the color ACS mosaic image in panel (b).  Panel (d) shows that the red stars are elliptically distributed with a major axis in the northeast-southwest direction.  The highest concentration of red stars are located in the northern part of field 2 (center field) and southern part of field 3 (upper field).  The blue stars (panel (c)) are scattered throughout the main body of the galaxy, ranging from low density in the Field 3 (lower field) to extremely prominent in the field 1.  The high density concentration of blue stars in the northern field are associated with an \hi\ supergiant shell \citep[SGS;][]{wal99a}.

The CMDs of each of the three fields are shown in Figure \ref{3cmd3}.  The RC, RGB, and AGB sequences of Fields 2 and 3 are virtually identical, a sign of the symmetric spatial distribution of red stars in these fields.  Field 1 contains less stars in these same regions, primarily because this field does not contain as large a section of the optical galaxy as the other two fields.  The MS and HeBs are most prominent in Field 3, which is expected due to the young stellar association near the SGS.    Fields 1 and 2 are mostly occupied by a low density of blue stars, with a few concentrated regions, which is reflected in the CMDs and blue density map.

The lifetime and recent SFHs of the three combined fields of IC 2574 are shown in Figures \ref{global_big} and \ref{recent_big}.  The oldest SFR is slightly higher than the lifetime average ($\sim$ 0.09 \msun\ yr$^{-1}$), while the intermediate age SFR is lower than the lifetime average ($f_{14}$ $=$ 0.65 and $f_{6}$ $=$ 0.20).  We see from both the lifetime and recent SFH plots that a large SF event occurred at $\sim$ 1 Gyr and continued to form stars at a rate above the lifetime SFR for $\sim$ 500 Myr ($b_{1G}$ $\sim$ 2 and $f_{1}$ $=$ 0.15).  Over the most recent 500 Myr, the SFR is still above the lifetime average ($b_{500}$ $\sim$ $b_{100}$ $\sim$ 1.4), but recently, IC 2574 is not forming stars as strongly as over the 500 Myr $-$ 1 Gyr period.

\section{Comparison of SFHs and SF Parameters}

In this section we examine the fraction of stars formed and birthrate parameters as both a function of luminosity and in comparison to the Local Group.  We divide the sample of galaxies based on their absolute blue magnitude with the dividing line at M$_{B}$ $=$ $-$15.0, an empirical separation used by \citet{lee07} based on \halpha\ EW measurements.  Additionally, we exclude both the Garland and \hoix\ from this analysis because they are likely to be stellar systems forming from tidal debris and not independent galaxies.  This leaves us with seven galaxies from the M81 sample for analysis.  While this sample is in the regime of small number statistics, we are able to compensate by probing SF at a variety of ages, something \halpha, UV, and IR flux surveys are unable to do.  Additionally, we are developing this analysis in anticipation of a larger number of galaxies with similar data quality (e.g., ANGST), where the number of galaxies will be a factor of ten larger.

Similar to \citet{lee07} and \citet{ken08} we present the mean and dispersion of the birthrate parameters as well as the fraction of stars formed  in two magnitude regimes, M$_{B}$ $>$ $-$15 and M$_{B}$ $\leq$ $-$15 (see Table \ref{tab4}).  Comparing the M81 Group galaxies in the two magnitudes regimes, we find that both types of galaxies formed comparable fractions of stars at ancient and intermediate times (see Table \ref{tab4}). The fraction of stars formed in the most recent Gyr is slightly higher in the more luminous galaxies ($\langle$$f_{1}$(M$_{B} \leq -$15)$\rangle$ $=$ 0.14 and $\langle$$f_{1}$(M$_{B}>-$15)$\rangle$ $=$ 0.08).  This difference may reflect  that the SFH correlates with mass, or that a higher fraction of recent SF boosts the luminosity of a galaxy increasing the probability that it will have M$_{B}$ $<$ $-$15.

Within the past Gyr, we can use the birthrate parameters to characterize the patterns of recent SF.   For the fainter dIs we find $\langle$$b_{500}$$\rangle$ and $\langle$$b_{1G}$$\rangle$ $\sim$ 1, consistent with the lifetime average SFR.  At more recent times, the SFR is elevated with $b_{100}$ $\sim$ 1.6.  Looking at the larger dIs, we find a similar increase of $b_{100}$ with $\langle$$b_{1G}$$\rangle$ and $\langle$$b_{500}$$\rangle$ $\sim$ 1.9 and $\langle$$b_{100}$$\rangle$ $\sim$ 2.3.  Despite the differences in the mean values of the birthrate parameters of the two classes of galaxies, the fractional increase in the birthrate parameter in the past 100 Myr is $\sim$ 1.5 for both types of galaxies.  Overall,  the fraction of stars formed and birthrate parameters are consistent with one another in the two luminosity regimes. 

Examining the dispersions in the birthrate parameters, we find that the values for $\sigma$$($$\log(b_{500})$$)$ and $\sigma$$($$\log(b_{1G})$$)$ are roughly equal for both types of galaxies, but over the past 100 Myr the dispersion in the birthrate parameters is $\sim$ 1.5 times greater for the brighter galaxies.  The increase in both dispersion and mean birthrate parameters in the past 100 Myr is not too surprising because the sample of dIs was selected to span a wide range of current SFRs as measured by \halpha\ emission \citep[Figure \ref{ha_ew}][]{lee07, ken08}.  However, the resultant dispersions are subject to our sample selection bias, and will be more meaningful with an expanded sample.

\subsection{Comparison with the Local Group}

An interesting application of measured SFHs is to compare the properties of galaxies in different groups.  The M81 Group is a more massive analog to the Local Group (LG) and experienced a major interaction within the last $\sim$ 300 Myr.  In the previous section, we explored the properties of the SFHs of the M81 Group dIs and found minor differences in the two luminosity regimes in both the fraction of star formed and birthrate parameters.  We apply the same statistics to select LG dIs with similar quality photometry and compare the results to those of the M81 Group dIs.  

To assemble a sample of LG galaxies, we selected those galaxies in the Local Group Photometry Archive \citep{had06} whose WFPC2 photometry is of comparable quality to our M81 Group dwarf observations.  The sample consists of Leo A, WLM, Sex A, IC 1613, NGC 6822, NGC 3109, and IC 10. The SFHs of both sets of galaxies were computed using the same method to allow for uniform comparison \citep{dol05}.  As we can see, the mean lifetime averaged SFR of the LG sample, $\sim$ 0.001 $M_{\odot}$  yr$^{-1}$,  is lower than that of the M81 Group sample,  $\sim$ 0.02 $M_{\odot}$  yr$^{-1}$.  Of course our selection of galaxies for comparison is based on photometric quality and not on physical characteristics and thus is subject to a selection bias.  For example, the samples are not comprable in their luminosity distributions.

\begin{figure}[t]
\begin{center}
\plotone{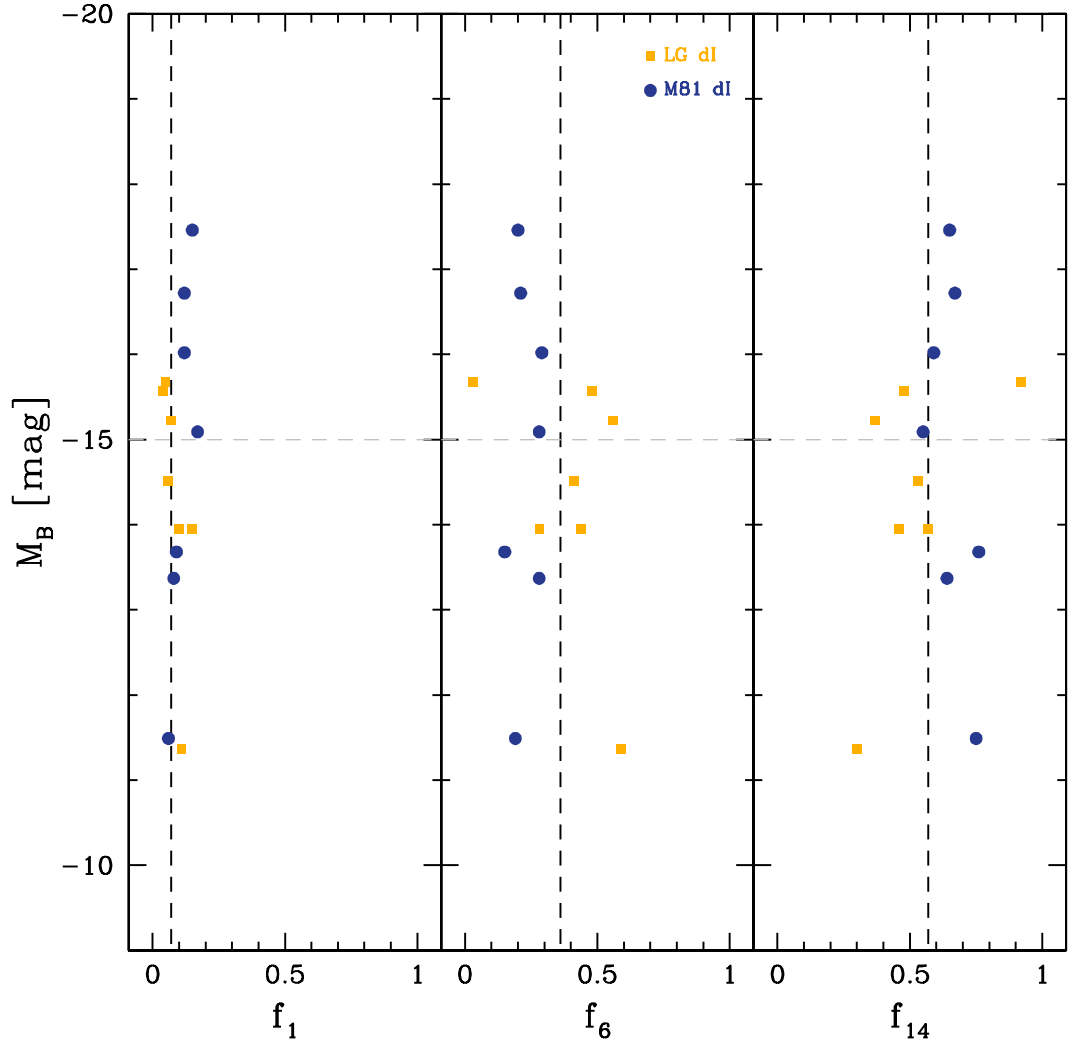}
\caption{The fraction of stars formed in the past 0 $-$ 1 Gyr ($f_{1}$), 1 $-$ 6 Gyr ($f_{6}$), and 6 $-$ 14 Gyr ($f_{14}$) plotted versus M$_{B}$ for both M81 Group (orange triangles) and Local Group (blue circles) galaxies.  The grey dashed line represents a transition luminosity noted by \citet{lee07} based on \halpha\ EW measurements.}  
\label{fm}
\end{center}
\end{figure}

Comparing the fraction of stars formed in M81 Group dIs and LG dIs, we find very little difference between the two groups (Figure \ref{fm} and Table \ref{tab4}).  While individual galaxies do not share the same $f$ values, the averages in the two luminosity regimes are very similar.  We extend this comparison by examining the average values of the composite sample with the expected values for a constant SFH (i.e., $f_{1}$ $=$ 0.07, $f_{6}$ $=$ 0.36, and $f_{14}$ $=$ 0.57).  While very few of the individual galaxies (Table \ref{tab3}) are consistent with the constant SFR $f$ values, the average of the composite sample, in both luminosity regimes, is in excellent agreement with a constant SFH.  It is important to note that this result does not imply that a constant SFH is a good model for any given dI.  Rather, this result emphasizes that dIs have SFHs that deviate randomly from a constant SFH model.  Additionally, we searched for a correlation between our $f$ parameters and $\Theta$, the tidal index \citep[see \S 6 and][]{kar04}, as well as our $f$ parameters and the center of mass of the respective groups, and found none.

\begin{figure}[t]
\begin{center}
\plotone{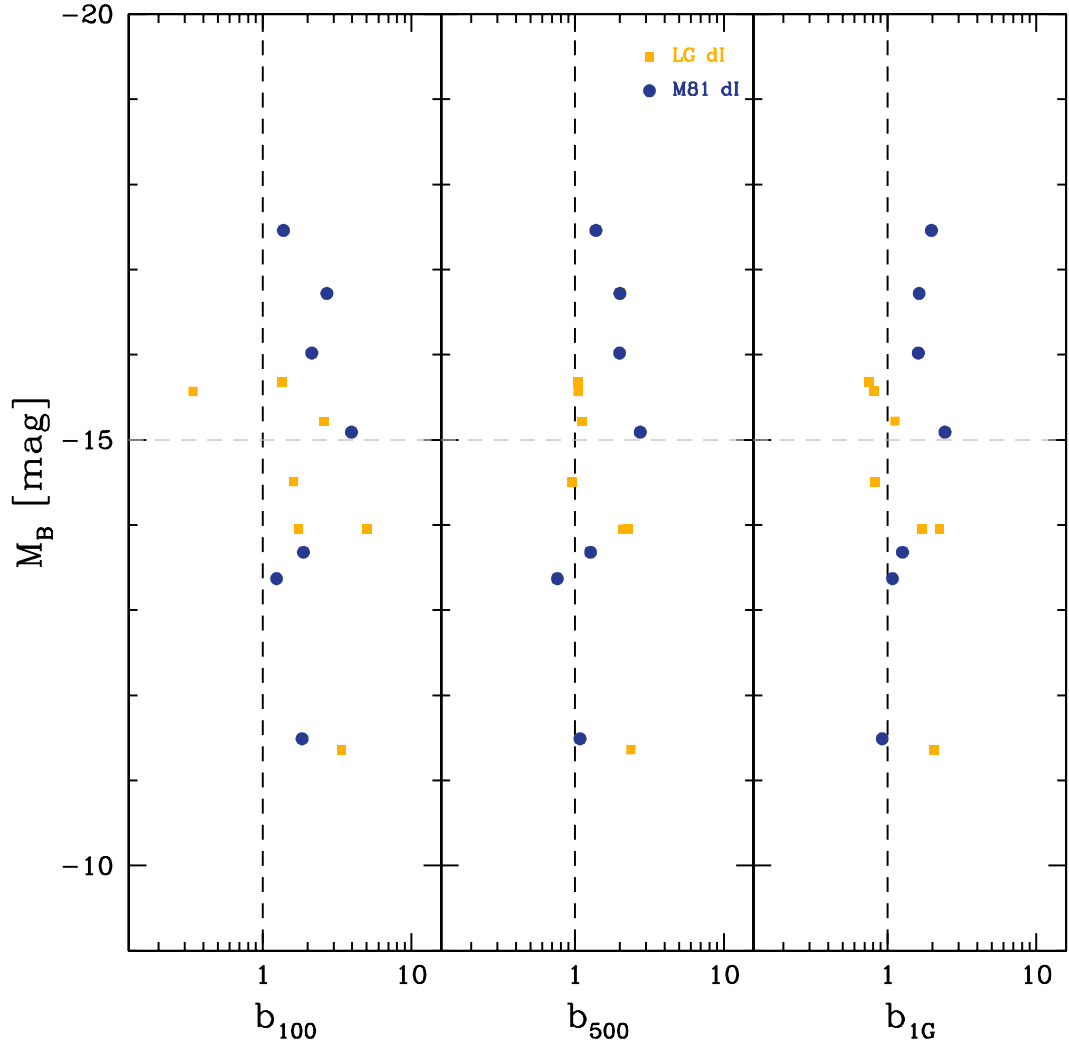}
\caption{The birthrate parameters plotted versus M$_{B}$ for both M81 Group (orange triangles) and Local Group (blue circles) galaxies.   The grey dashed line represents a transition luminosity noted by \citet{lee07} based on \halpha\ EW measurements.}  
\label{bm}
\end{center}
\end{figure}

Similarly, we compare the birthrate values for the M81 dIs and LG dIs, finding little difference between the composite values in the two luminosity regimes (Figure \ref{bm} and Table \ref{tab4}).  On average the the faint LG galaxies have higher mean birthrate parameters than their M81 counterparts.  Conversely, the bright M81 Group dIs have higher mean birthrate parameters than the bright LG dIs.  Examining the composite values, we find that they are consistent with one another and are well above what we would expect for a constant SFH (i.e., $b$ = 1).  This conclusion suggests that the birthrate parameters for dIs are not correlated with luminosity and are randomly distributed.  However, we need to examine a larger sample to support this conclusion.  We also examined trends between our $b$ parameters and $\Theta$ and between our $b$ parameters and the center of mass of the respective groups, but found no clear correlation.

\section{Formation and Evolution of Small and Potential Tidal Dwarfs}

Of the 4 faintest galaxies, there is evidence that the Garland and \hoix\ are tidal dwarfs \citep{yun99, mak02}. There have also been suggestions that  DDO 53 is the result of a merger \citep{beg06}, and that M81 Dwarf~A may be of tidal origin \citep{bur04}.   All of these galaxies share a common feature an of increased SFR over the past 100 Myr,  which may be due to induced SF as a result recent interactions.  We  now assess the evidence to determine which of these galaxies may be recently formed tidal dwarfs and which may be genuinely old, but with recent SFR enhancements.

\begin{figure}[t]

\plotone{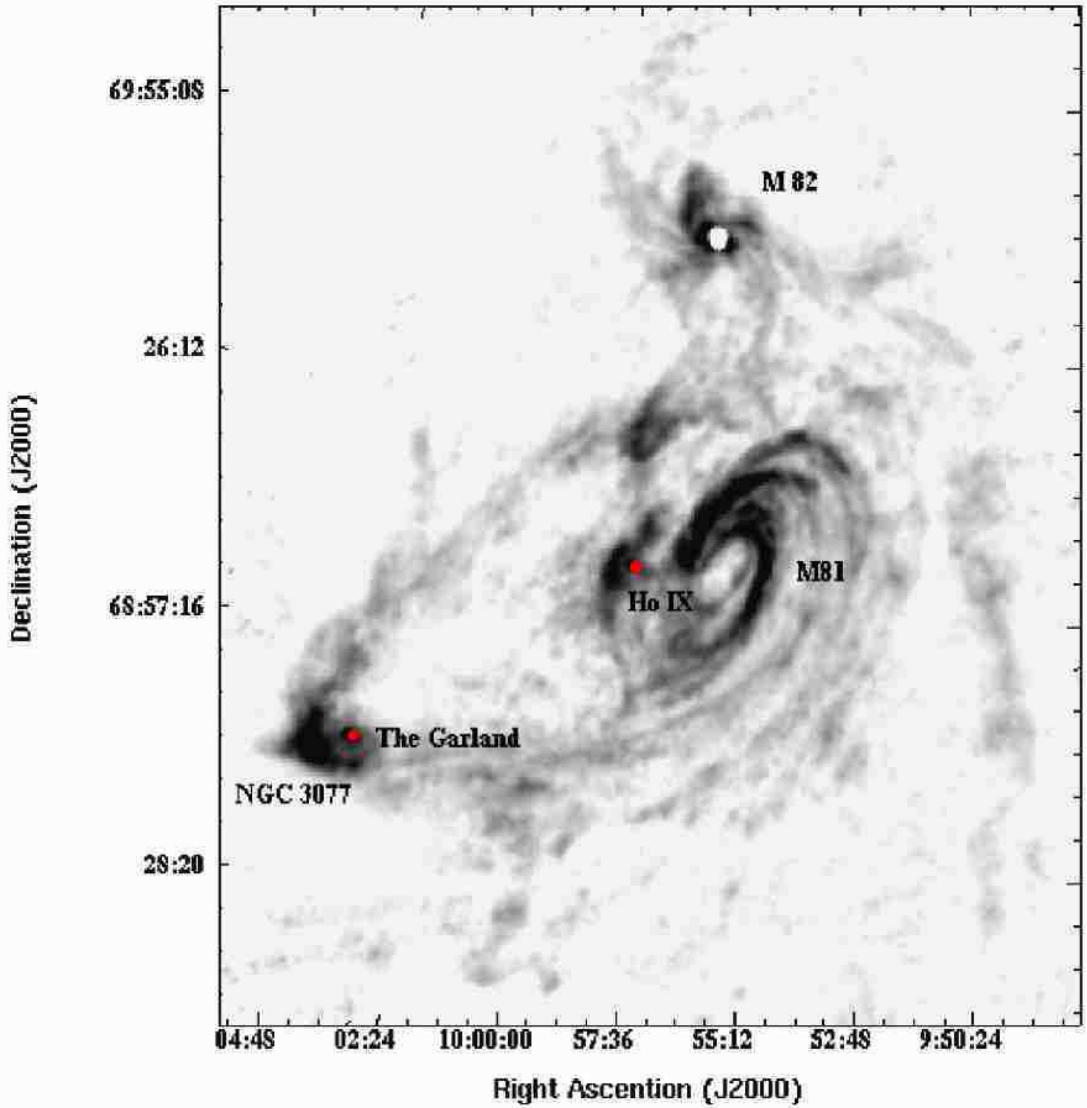}
\caption{A high resolution \hi\ image of the M81 Group \citep{yun94} with the locations of the three larger galaxies (M81, M82, and NGC 3077) and two candidate tidal dwarfs (the Garland and \hoix) indicated.}  
\label{hi_map}

\end{figure}

Close interactions between galaxies are capable of leaving tidal debris that may be the raw material for the formation of new stellar systems, perhaps even new independent galaxies \citep[e.g.,][]{zwi56, tom72, duc97}.  In the M81 Group, \hi\ observations \citep{yun94} reveal evidence of tidal streams and debris indicative of a recent interaction of larger galaxies within the group.  Subsequent modeling of the group dynamics  \citep{yun99} find the three large galaxies (M81, M82, and NGC 3077) had an interaction $\sim$ 250 Myr for M81 and NGC 3077 and $\sim$ 220 Myr ago for M81 and M82.  Within this tidal debris around NGC 3077, we find the Garland, an overdensity of young stars (see Figure \ref{hi_map}), also coincident with peaks in \hi\ and CO \citep{van79, wal99b, yun94}.  Based on our SFH, we find a rise in the SFR at $\sim$ 250 Myr, slightly more recently than the proposed interaction timescale.  Similar to the Garland, \hoix\ is located in the inner tidal arm of M81 (see Figure \ref{hi_map}), and also experienced a rise in SFR, $b_{100}$ $\sim$ 20, slightly more recently than the interaction time of 220 Myr.  Further, these two systems have large tidal indices, $\Theta_{Garland}$ $=$ 4.0 and $\Theta_{\hoix}$ $=$ 3.3, which indicate the relative amount of gravitational disturbance from a neighboring galaxy \citep[see Equation 8 in][]{kar04}. Large positive values indicate a high degree of disturbance and more negative numbers imply a larger degree of isolation. 

While it is difficult to determine if the Garland and \hoix\ are tidal dwarfs, it is reasonably straightforward to see that the Garland and \hoix\ are not dIs.  Comparing red star distributions with the two faint dIs in the sample, we see that both DDO 53 and M81 Dwarf~A have prominent red star components that are distinct from the background and coincident with the high density of blue stars.  However, the Garland and \hoix\ \citep{sab08} do not show any evidence of a red star overdensity coincident with the blue stars.  From this we can conclude that the Garland and \hoix\ are not dIs.

Beyond comparing the Garland and \hoix\ to dIs, we cannot distinguish if the Garland and \hoix\ are true tidal dwarfs or alternatively stellar systems forming in their larger companion.  A primary criteria for determination of a tidal dwarf is a low mass-to-light ratio, due to the absence of dark matter \citep[e.g.,][]{bar92, duc00}, which we cannot measure from the resolved stellar component.  Without this measurement, we can only conclude that these are possible tidal dwarf galaxies.

\hi\ observations and analysis by \citet{bur04} suggest that M81 Dwarf~A may also be of tidal origin formed from tidal debris originally belonging to \hoii. However, M81 Dwarf~A shows no dramatic increase in the recent SFH and the older stellar population is centrally concentrated, both unlike the candidate tidal dwarfs in our sample.  Examining the SFH, neither the ancient nor recent SFHs show evidence of tidal formation.  M81 Dwarf~A formed 75\% of its stars 6 $-$ 14 Gyr ago, which is unlike \hoix.  The Garland does show a higher fraction of stars formed in the same bin, but this is likely due to red star contamination from NGC 3077.  At recent times, M81 Dwarf~A does show an increasing trend in birthrate parameters toward the present, but there is not a sudden increase in the SFR as is the case with the other candidate tidal dwarf galaxies in the M81 Group sample.  Spatially, the red star distribution is fairly symmetric, which is not like that of the other tidal dwarfs.  Similarly, the \hi\ component of M81 Dwarf~A does not show an asymmetry toward or any connection to a nearby companion \citep{beg06, wal08}. In agreement with these observations, M81 Dwarf~A has  a tidal index of 0.7 (Table \ref{tab3}), which indicates it is fairly isolated.  Thus, M81 Dwarf~A is not a tidal dwarf candidate like the Garland or \hoix.  

The unusual kinematics of the \hi\ in DDO 53 led \citet{beg06} to speculate that DDO 53 may be a product of a recent merger between two fainter dwarfs.  The \hi\ kinematics are comparable to other galaxies thought to be late-stage mergers \citep{hun02, kor03}.  From our spatial analysis (Figure \ref{ddo53_image}), we do find that the red stars are not symmetric, which could be due to a merger or at least gravitational influence of another galaxy.  The blue stars are centrally concentrated and align well with peaks in both \halpha\ and \hi\ observations \citep{wal08, beg06}, but they do not follow the distribution of red stars, which we might expect if there was a recent gravitational disturbance.  The recent SFH of DDO 53 reveals that it is forming stars at a higher rate at the present than any other time within the more recent 1 Gyr.  Outside of this dramatic increase, DDO~53 has been forming stars at or below the lifetime average over the past 1 Gyr, which would not be consistent with a merger in this timeframe.  Similarly, with a tidal index of 0.7, DDO~53 is a relatively isolated galaxy and does not appear to have been gravitationally disturbed recently.  It is possible that DDO~53 is the result of a merger, but such an interaction would have had to take place more than 1 Gyr ago according to our observations and analysis.  

We can also compare the the chemical evolution measured from the SFH fitting method to search for differences between the small and candidate tidal dwarfs.  M81 Dwarf~A and DDO 53 show very little evolution in their chemical abundances over their lifetimes (see Table \ref{tab3}) and DDO 53 has a 1 Gyr value close to the observed abundance (Tables \ref{tab1} and \ref{tab3}).  The galaxies underwent very little in the way of chemical evolution over their lifetimes, which would be consistent with isolated, i.e., non-interacting galaxies. In contrast, the candidate tidal dwarfs have significant chemical evolution \citep[e.g.,][]{bar92}.  As previously mentioned, for the candidate tidal dwarfs we did not require the chemical evolution to monotonically increase.  The motivation for this was to help find evidence if there truly were two populations of stars in these fields.  The difference in the ancient and recent metallicity values (0.6 dex for the Garland and 0.3 dex for \hoix) and the fact that the younger components are relatively metal poor are both signs, albeit indirect, that there are mixed populations in those fields; the more metal rich stars from the large companion and the more metal poor stars formed from tidal debris.

\section{Conclusions}

From HST/ACS imaging we use resolved stellar populations to study the temporal and spatial properties of nine diverse dIs in the M81 Group.  Photometry from the HST/ACS images was used to create CMDs and trace the temporal evolution of each galaxy via measured SFHs.  From the CMD, we isolated red (old) and blue (young) stars and created stellar density maps to better understand the stellar spatial components.  Notably, we analyze two candidate tidal dwarf galaxies, the Garland and \hoix, and find that they are clearly different from any of the other galaxies in the sample, both in their SFHs and spatial stellar distributions, despite having similar luminosities.  They are particularly interesting because they are two of the nearest candidate tidal dwarf galaxies in the universe, as we do not have similar galaxies in the Local Group.  

In addition to analyzing the galaxies on an individual basis, we divided the sample into two groups, based on an empirical separation in luminosity of star forming galaxies in the Local Volume \citep{lee07, ken08}.  Based on this division, we compute the fraction of stars formed and birthrate parameters, and compare these values between the two regimes, finding very little difference.  We extend our comparison of dIs to the LG, selecting LG galaxies with similar quality photometry for comparison.   Again, comparing the M81 Group dIs and LG dIs, we find very little difference in their SF characteristics (i.e., fraction of stars formed and birthrate parameters).  Analyzing the composite population of the M81 and LG dIs, we find that the average values of the fraction of stars formed in both luminosity regimes are consistent with that of a constant SFH.   Importantly, we note that individual galaxies are not well modeled by a constant SFH.  We interpret this result to mean that the fraction of stars formed in each galaxy is essentially random (or stochastic), but that with a larger sample, the mean values converge on those consistent with a constant SFH.  We searched for trends in the birthrate parameters as a function of luminosity, but found none, supporting the idea that stochastic SF is important in dIs.  With our future work, we will explore the same type of analysis with a larger sample, as well as comparing spatially resolved SFHs to multi wavelength observations to give us further insight into the impact of SF on the evolution of low mass galaxies.

\section{Acknowledgments}

Support for this work was provided by NASA through grants GO-9755 and GO-10605
from the Space Telescope Science Institute, which is operated by
AURA, Inc., under NASA contract NAS5-26555. DRW is grateful for support 
from a Penrose Fellowship. We would like to thank the anonymous referee for prompt and valuable comments. DRW would also like to thank Julianne Dalcanton for her insightful comments and discussion.  EDS is grateful for partial support from the University of Minnesota.
This research has made use of NASA's Astrophysics Data System
Bibliographic Services and the NASA/IPAC Extragalactic Database
(NED), which is operated by the Jet Propulsion Laboratory, California
Institute of Technology, under contract with the National Aeronautics
and Space Administration.

\clearpage

\begin{landscape}
\begin{deluxetable}{lccccccccc}
\tablecolumns{10}
\tabletypesize{\footnotesize}
\small
\tablewidth{0pt}
\tablehead{
    \colhead{Galaxy} &
    \colhead{Alternate} &
    \colhead{$M_{B}$} &
       \colhead{$Log(M_{\hi})$\tablenotemark{1}} &
      \colhead{$(m-M)_{0}$} &
      \colhead{$A_{V}$\tablenotemark{5}}  &
      \colhead{Z} &
      \colhead{Field} &
      \colhead{RA} &
    \colhead{Dec} \\
}
\tablecaption{Basic Properties of M81 Group Dwarf Galaxies}
\startdata                                                              
Garland &  & $-$11.40 & 7.54 &   27.75\tablenotemark{4} & 0.22   & ... & & 10:03:51.70 & +68:41:25 \\
Dwarf~A & K52 & $-$11.49 & 7.12 & 27.75\tablenotemark{2} & 0.07 &  ... &   &    08:23:56.00   &  +71:01:45 \\
DDO 53 & UGC 4459, VII Zw 238 & $-$13.37 & 7.61 & 27.76\tablenotemark{2} & 0.12 &   $-$1.04\tablenotemark{9} & &        08:34:07.20 & +66:10:54  \\
\hoix\  & UGC 5336, DDO66, K62 & $-$13.68 & 8.50 & 27.84\tablenotemark{3}   & 0.26  &   $-$0.42\tablenotemark{10, 11} & &       09:57:32.00 &  +69:02:45 \\
\hoi\   & UGC 5139, DDO63 &  $-$14.49 & 8.12 & 27.92\tablenotemark{2} & 0.16  &   $-$0.96\tablenotemark{6} & &     09:40:32.10  &   +71:11:12  \\
DDO165 & UGC 8201 &  $-$15.09 &  8.14 & 28.30\tablenotemark{2} & 0.08 &   ... &  &   13:06:26.40 & +67:42:24  \\
NGC 2366 & DDO42 & $-$16.02 & 8.85 & 27.52\tablenotemark{2} & 0.12 &  $-$0.75\tablenotemark{7} &      &  &   \\
 & & && & & &  1 & 07:28:43.96 & +70:42:04\\
 & & && & &  & 2 & 07:28:59.60 & +69:14:15\\

 \hoii\  	    & UGC 4305, DDO50 & $-$16.72 & 8.99 & 27.65\tablenotemark{2} & 0.11  &   $-$0.74\tablenotemark{7,8} &     &   &   \\
  & & &&  & &  & 1 & 08:18:58.96 & +70:42:04\\
    & & &&  & & &  2 & 08:19:20.50 & +70:43:40\\
IC 2574	    & DDO81, UGC 5556 & $-$17.46 & 9.23 &   28.02\tablenotemark{2} & 0.12  &   $-$0.57\tablenotemark{6,7} & & & \\
 & & && & & & 1 & 10:28:43.14 & +68:27:04\\
 & & && & & & 2 & 10:28:23.07 & +68:24:36\\
 & & & &  & & &  3 & 10:27:50.00 & +68:22:55 \\

\enddata\\
\tablecomments{ (1) \cite{kar07} (2)\cite{kar02} (3) \cite{kar06} (4) CMD fitting program \citep{dol02} (5) \cite{sch98} (6) \cite{mil96} (7) \cite{mas91} (8) \cite{lee03} (9) \cite{ski89} (10) \cite{van98} (11) \cite{den02}}
\label{tab1}
\end{deluxetable}
\end{landscape}
\clearpage

\begin{deluxetable}{lcccccc}
\tablecolumns{7}
\tabletypesize{\footnotesize}
\small
\tablewidth{0pt}
\tablecaption{Observations and Completeness}
\tablehead{
    \colhead{Galaxy} &
     \colhead{Field} &
      \colhead{Int. time} &
      \colhead{Int. time} &
      \colhead{No. of Stars}   &
      \colhead{50\% completeness} &
       \colhead{50\% completeness} 
      \\
       \colhead{} &
     \colhead{} &
      \colhead{F555W (secs)} &
      \colhead{F814W (secs)}  &
       \colhead{in CMD} &
      \colhead{F555W} &
       \colhead{F814W}  \\
      
}
\startdata
Garland\tablenotemark{1} &  & 9600 & 19200 & 127,778 & 28.6 & 28.1 \\
Dwarf~A & &5914 & 5936 & 17,450 & 28.4 & 27.8 \\
DDO 53 & & 4768 & 4768 & 67,903 & 28.1 & 27.8  \\
\hoix\  & & 4768 & 4768 & 68,373 & 28.3 & 27.7 \\
\hoi\   & & 4446 & 5936 & 121,198 & 28.0 & 27.8 \\
DDO 165 & & 4768 & 4768 & 120,281 & 28.0 & 27.2 \\
NGC 2366 & 1 & 4780  & 4780 & 246,750 &  27.8 & 27.4 \\
 & 2 & 4780 & 4780 & 232,569 & 27.9 & 27.4 \\
\hoii\   & 1 & 4660 & 4660 & 239,742  & 27.8 & 27.3 \\
 & 2 & 4660 & 4660 & 226,602  & 27.9 & 27.5\\
IC 2574\tablenotemark{2} & 1 & 6400 & 6400 & 253,736 & 28.2 & 27.7\\
 & 2 & 4784 & 4784 & 234,369 & 27.7 & 27.0\\
 & 3 & 4784 & 4784 & 159,169 & 28.1 & 27.6\\
\enddata
\tablecomments{All data observed under HST Program GO-10605 (PI Skillman), except (1)  GO-9381 (PI Walter) and (2) GO-9755 (PI Walter).  }
\label{tab2}
\end{deluxetable}

\clearpage

\begin{landscape}
\begin{deluxetable}{lcccccccccccc}
\tablecolumns{12}
\small
\tablewidth{0pt}
\footnotesize
\tablecaption{SF Properties of M81 Group and Local Group Dwarf Galaxies }
\tablehead{
    \colhead{Galaxy} &
    \colhead{Global Average SFR} &
     \colhead{$b_{100}$} &
      \colhead{$b_{500}$} &
      \colhead{$b_{1G}$} &
      \colhead{$f_{1G}$} &
      \colhead{$f_{6G}$} &
      \colhead{$f_{14G}$} &
      \colhead{ $[$$\frac{M}{H}$$]_{1G}$} &
      \colhead{ $[$$\frac{M}{H}$$]_{6G}$} &
      \colhead{ $[$$\frac{M}{H}$$]_{14G}$} &
      \colhead{ $\Theta$ } \\
      
        \colhead{} &
    \colhead{10$^{-2}$ ($M_{\odot}  yr^{-1}$) } &
     \colhead{} &
      \colhead{} &
      \colhead{} &
       \colhead{} &
      \colhead{} &
      \colhead{} &
           \colhead{} &
         \colhead{} &
         \colhead{} &
      \colhead{}   \\
}
\startdata

Garland & 0.61 & 0.37 & 0.25 & 0.15  &  0.01 & 0.09 & 0.91 &  $-$1.0 & $-$0.3 & $-$0.4 & 4.0 \\
Dwarf~A & 0.10 & 1.84 & 1.08 & 0.92 &   0.06 & 0.19 &0.75 & $-$1.3 & $-$1.3 & $-$1.6 & 0.7\\
DDO 53 & 0.52 & 1.24 & 0.76 & 1.08 &   0.08 & 0.28 & 0.64 &  $-$1.2 & $-$1.2 & $-$1.2 & 0.7\\
\hoix~ & 0.19 & 19.83 & 5.46 & 4.21 &  0.30 & 0.46 & 0.24 & $-$1.4 & $-$0.8 & $-$1.1 & 3.3\\
\hoi~ & 1.03 &1.88 & 1.27 & 1.26 &   0.09 & 0.15 & 0.76 & $-$1.2 & $-$1.3 & $-$1.4 & 1.5\\
DDO 165 & 1.27 & 3.95 & 2.74 & 2.43 &  0.17 & 0.28 & 0.55 &   $-$1.3 & $-$1.4 & $-$1.7 & 0.0\\
NGC 2366 & 3.37 & 2.14 & 1.99 & 1.61 &   0.12 & 0.29 & 0.59 & $-$1.2 & $-$1.3 & $-$1.4 & 1.0\\
\hoii~ & 3.64 & 2.70 & 2.00 & 1.63 & 0.12 & 0.21 & 0.67 &  $-$1.3 & $-$1.3 & $-$1.4 & 0.6\\
IC 2574 & 8.63 & 1.38 & 1.38 & 1.97  &   0.15 & 0.20 & 0.65 & $-$1.0 & $-$1.1 & $-$1.2 & 0.9\\
\hline
\hline
Leo A & 0.02 & 3.39 & 2.37 & 2.05 &  0.11 &  0.59 & 0.30 &   $-$1.3 & $-$1.4 & $-$1.4 & 0.2\\ 
WLM & 0.08 & 1.73 & 2.08 & 2.24 &  0.10 & 0.44 & 0.46 &  $-$1.0 & $-$1.0 & $-$1.3 & 0.3\\
Sex A & 0.18 & 5.06 & 2.28 & 1.70 &   0.15 & 0.28 & 0.57 &  $-$1.4 & $-$1.4 & $-$1.6 & $-$0.6\\
IC 1613 & 0.10 & 1.61 & 0.95 & 0.83 &   0.06 & 0.41 & 0.53 & $-$0.7 & $-$1.0 & $-$1.7 & 0.9\\
NGC 6822 & 0.17 & 2.59 & 1.12 & 1.12 &   0.07 & 0.56 & 0.37 & $-$0.5 & $-$0.8 & $-$1.9 & 0.6 \\
NGC 3109 & 0.56 & 1.36 & 1.05 & 0.75 &  0.05 & 0.03 & 0.92 &  $-$0.8 & $-$0.9 & $-$1.2 & $-$0.1\\
IC 10 & 0.87 & 0.34 & 1.05 & 0.81&   0.04 & 0.48 &  0.48 & $-$0.4 & $-$0.5 & $-$0.8 & 1.8\\

\enddata
\tablecomments{\scriptsize{$b_{100}$: SFR over the last 100 Myr normalized to the lifetime average SFR, $b_{500}$: SFR over the last 500 Myr normalized to the lifetime average SFR, $b_{1G}$: SFR over the last 1 Gyr normalized to the lifetime average SFR.  $f_{1}$ is the fraction of stars formed from 0 $-$ 1 Gyrs, $f_{6}$ from 0 $-$ 6 Gyr, and $f_{14}$ from 6 $-$ 14 Gyr.  $[$$\frac{M}{H}$$]_{1G}$ is the mean metallicity from 0 $-$ 1 Gyr, $[$$\frac{M}{H}$$]_{6G}$ from 1 $-$ 6 Gyr, and $[$$\frac{M}{H}$$]_{14G}$ is  from 6 $-$ 14 Gyr. Note that for a constant SFR , $f_{1}$ $=$ 0.07, $f_{6}$ $=$ 0.36, and $f_{14}$ $=$ 0.57.  All metallicity values are from our CMD fitting program. $\Theta$ is the tidal index of the galaxy as calculated by \citet{kar04, kar07}, with higher values indicating a higher likelihood of gravitationally induced tidal interactions.  Errors on the $f$ and $b$ values are dominated by systematic errors, which are very small, and do not noticeably affect the values in the table, and are not listed.}}
\label{tab3}
\end{deluxetable}
\end{landscape}

\clearpage

\begin{deluxetable}{cccccccccc}

\small
\tablewidth{0pt}
\footnotesize
\tablecaption{Star Formation Statistics}
\tablehead{

&&\multicolumn{2}{c}{M81 Group} & & \multicolumn{2}{c}{Local Group} & &  \multicolumn{2}{c}{M81 + LG}\\
\cline{3-4}\cline{6-7}\cline{9-10}
\\[-1ex]
& \colhead{M$_{B}$} & \colhead{$> -15$} &  \colhead{$\leq -15$} & & \colhead{$> -15$} &  \colhead{$\leq -15$} & & \colhead{$> -15$} &  \colhead{$\leq -15$}
}

\startdata
$\langle$$f_{1}$$\rangle$ & & 0.08 & 0.14 & & 0.11 & 0.05 & & 0.09 & 0.10\\
$\langle$$f_{6}$$\rangle$ & & 0.21 & 0.25 & & 0.43 & 0.36 & & 0.34 & 0.29\\
$\langle$$f_{14}$$\rangle$ & & 0.71 & 0.61 & & 0.46 & 0.59 & & 0.57 & 0.61\\
$\sigma$$($$f_{1}$)$$& & 0.02 & 0.02 & & 0.04 & 0.02 & & 0.03 & 0.05\\
$\sigma$$($$f_{6}$)$$& & 0.07 & 0.05 & & 0.12 & 0.29 & & 0.15 & 0.18\\
$\sigma$$($$f_{14}$)$$& & 0.07 & 0.06 & & 0.12 & 0.29 & & 0.16 & 0.17\\
$\langle$$\log(b_{100})$$\rangle$ & &  0.21 & 0.37 & & 0.42 & 0.03 & & 0.33 & 0.23  \\
$\langle$$\log(b_{500})$$\rangle$ & &  0.01 & 0.29 & & 0.26 & 0.03 & & 0.15 & 0.18  \\
$\langle$$\log(b_{1G})$$\rangle$ & &  0.03 & 0.27 & & 0.20 & $-$0.06 & & 0.13 & 0.13  \\
$\sigma$$($$\log(b_{100})$$)$ & & 0.10 & 0.19 & & 0.24 & 0.45 & & 0.21 & 0.35\\
$\sigma$$($$\log(b_{500})$$)$ & & 0.11 & 0.12 & & 0.19 & 0.02 & & 0.20 & 0.17\\
$\sigma$$($$\log(b_{1G})$$)$ & & 0.07 & 0.08 & & 0.20 & 0.09 & & 0.17 & 0.19\\

 \enddata

\tablecomments{ The mean and dispersion of the fraction of stars formed and the logarithm of the birthrate parameters for the M81 Group, Local Group, and combined set of galaxies.  We have divided the sample at M$_{B}$ $=$ $-$15 based on a similar division  by \citet{lee07}. Dispersions indicated the range spanned by the sample, but do not represent a true dispersion intrinsic to the population, due to sample selection biases.  Note that for a constant SFR , $f_{1}$ $=$ 0.07, $f_{6}$ $=$ 0.36, and $f_{14}$ $=$ 0.57.  Errors on the $f$ and $b$ values are dominated by systematic errors, which are very small, and do not noticeably affect the values in the table, and are not listed. }
\label{tab4}
\end{deluxetable}

\newpage
\clearpage

\newpage

\newpage

\newpage

\newpage

\newpage

\newpage

\newpage

\newpage

\newpage

\newpage

\newpage

\newpage

\newpage

\newpage

\newpage
\clearpage

\newpage

\newpage
\clearpage

\newpage
\clearpage

\newpage
\clearpage

\clearpage

\end{document}